\documentclass{article}

\usepackage{PRIMEarxiv}

\usepackage{a4}
\usepackage[utf8]{inputenc} % allow utf-8 input
\usepackage[T1]{fontenc}    % use 8-bit T1 fonts
\usepackage{hyperref}       % hyperlinks
\usepackage{url}            % simple URL typesetting
\usepackage{booktabs}       % professional-quality tables
\usepackage{amsfonts}       % blackboard math symbols
\usepackage{nicefrac}       % compact symbols for 1/2, etc.
\usepackage{microtype}      % microtypography
\usepackage{lipsum}
\usepackage{fancyhdr}       % header
\usepackage{graphicx}       % graphics
\usepackage{caption}
\usepackage{xcolor}
\usepackage{subfigure}
\usepackage[labelfont=bf]{caption}
\usepackage[tableposition=top]{caption}
\graphicspath{{media/}}     % organize your images and other figures under media/ folder
\usepackage{array}
\usepackage{multirow}

\usepackage[sort&compress, numbers]{natbib}
\title{Dynamic and Context-Dependent Stock Price Prediction \\
Using Attention Modules and News Sentiment
}

\author{ {\hspace{1mm}Nicole Königstein}\thanks{ impactvise AG,  nicole.koenigstein@impactvise.com} \\}

\begin{document}
\maketitle

\begin{abstract}
The growth of machine-readable data in finance, such as alternative data, requires new modeling techniques that can handle non-stationary and non-parametric data. Due to the underlying causal dependence and the size and complexity of the data, we propose a new modeling approach for financial time series data, the $\alpha_{t}$-RIM (recurrent independent mechanism). This architecture makes use of key-value attention to integrate top-down and bottom-up information in a context-dependent and dynamic way. To model the data in such a dynamic manner, the $\alpha_{t}$-RIM utilizes an exponentially smoothed recurrent neural network, which can model non-stationary times series data, combined with a modular and independent recurrent structure. We apply our approach to the closing prices of three selected stocks of the S\&P 500 universe as well as their news sentiment score. The results suggest that the $\alpha_{t}$-RIM is capable of reflecting the causal structure between stock prices and news sentiment, as well as the seasonality and trends. Consequently, this modeling approach markedly improves the generalization performance, that is, the prediction of unseen data, and outperforms state-of-the-art networks such as long short-term memory models.
\end{abstract}

% keywords can be removed

\section{Introduction}
Non-stationary time series data is common in finance. In the case of price series, each value depends on a long history of prior levels. However, most machine learning models that predict financial time series expect stationary inputs. Consequently, these models rely on standard stationarity transformations, such as integer differentiation, to produce returns, which have a memory cut-off hence the series loses important signals \citep{prado_2018}, \citep{sugiyama_kawanabe_2012}. 
In addition, in the field of causal inference, the concept of independent or autonomous processes has proven to be important because these processes make the model capable of cause and effect inference \citep{causalityPearl}, \citep{DBLP:journals/corr/abs-1709-08568}. Thus, a complex model may be thought of as a collection of separate processes or “causal” modules. As a result, individual modules may be robust or invariant even when other modules change, as in a distribution shift \citep{peters_janzing}, \citep{schoelkopf2012causal}. Furthermore, most machine learning modules are monolithic and based on bottom-up signals, i.e. directly observed content, in contrast to a top-down signal, which is based on past experience and short-term memory. Moreover, human cognition has a modular structure with sparse interactions, as described by Carruthers in his book \textit{The Architecture of the Mind} \cite{Carruthers2006-CARTAO-26}. Carruthers argues that one of the distinctive characteristics of the human mind is that it is composed of numerous cognitive systems, each of which communicates with a small number of other systems or experts, each of which has little influence over the processes occurring within them. Thus, the human mind is flexible and capable of practical reasoning and thus gains the capacity for scientific thinking. Consequently, if we think of the brain as capable of solving problems by using different systems (or modules), we hypothesize that it could be beneficial to leverage this kind of structure by learning separate processes that can be reused, constructed, and flexibly re-purposed. Humans also seldom utilize all available inputs to complete tasks. For these reasons, using sparse interactions and focusing attention in machine learning models may reduce learning difficulties by minimizing interference. In other words, models that learn in this manner may more accurately reflect the causal structure of the data and hence generalize more effectively \citep{Simon1991}, \citep{peters_janzing}. Accordingly, within this context, we analyze how such a modeling approach can be used to incorporate stock prices and news sentiment to predict stock price movements.

\section{Related Work}
\label{sec:headings}

Prior to the use of deep learning within the field of financial times series prediction, methods such as ARIMA and its modifications were mainly used. For instance, Minyoung Kim replaced the conventional maximum likelihood estimation with an asymmetric loss function owing to the asymmetric distribution of financial time series returns \citep{ARIMAnew}, and Adebiyi et al. evaluated the ARIMA and artificial neural network prediction performance on NYSE stocks (e. g., stock price of Dell Incorporation)  \cite{articleARIMAvsNN}. The empirical findings of these studies demonstrated the superiority of neural networks over ARIMA models. Recent advances in deep learning have resulted in enhanced model performance, resulting from the increased computing capacity and the ability to understand non-linear connections embedded in a variety of financial variables. \\

Sreelekshmy Selvin et al. examined three distinct deep learning architectures for stock prediction, including recurrent neural network (RNN), long short-term memory (LSTM), and convolutional neural network (CNN)-sliding window models for the prediction of Nation Stock Exchange of India listed stocks \citep{Selvin2017StockPP}. Within their study, they concluded that the CNN outperforms the RNN and LSTM because it is capable of identifying trend changes in the stocks. Further, Panday et al. evaluated a hybrid algorithm, which uses an LSTM to predict stock values and a sentiment analysis to verify the predictions \citep{paperSentiment}. Their findings indicated that there is, in fact, a relationship between public sentiment and stock prices. Moreover, Matthew Dixon has introduced a new class of smoothed RNNs, the $\alpha$-RNN and the $\alpha_{t}$-RNN \citep{dixon2020alpha} , and shown that they are superior and more robust than simple RNNs and ARIMA models for industrial forecasting. Recently, the concept of attention has also acquired considerable importance in the area of artificial intelligence as an essential component of neural networks for a large number of Computer Vision applications \citep{attentionMultimedia} \citep{wang-etal-2016-attention}, speech recognition, natural language processing \citep{Galassi_2020} and, more recently, for financial time series prediction  \citep{attFinancial1} \citep{attFinancial2}. In those cases, the experimental findings demonstrated that the networks with attention outperformed models such as ARIMA and LSTM without attention in financial time series prediction.

\section{Proposed Model}

In this section, we outline our proposed approach, which relies on two deep learning architectures, namely the recurrent independent mechanism  \citep{goyal2019recurrent} and a general class of exponentially smoothed RNNs, the $\alpha$-RNN and the $\alpha_{t}$-RNN \citep{dixon2020alpha}. More specifically, we use the traditional RIM architecture, where the modules are based upon an $\alpha_{t}$-RNN cell as a recurrent model, as opposed to an LSTM cell.

\subsection{Recurrent Independent Mechanism} The RIM is a recurrent architecture with multiple modules of recurrent cells, which operate with independent transition dynamics  \citep{henaff2017tracking}, \citep{kipf2018neural}. They communicate sparsely via attention and compete for the most relevant input for each time step. In RIMs, the updates for the hidden state follow a three-step process: 

\begin{enumerate}
\item A subset of modules is selectively activated based on the determination of the relevance of the modules input. 
\item The activated modules independently process the information made available to them. 
\item The RIMs communicate with one another sparingly via key-value attention, then the active modules gather contextual information from all the other modules and consolidate this information in their hidden state. 
\end{enumerate}

The entire model is subdivided into $k$ small subsystems, so-called RIMs, in which each represents a recurrent model to capture the dynamics in observable sequences. As a consequence, each RIM has its own unique functions that are automatically trained from data, resulting in the vector-valued state $h_{t,k}$ at timestep $t$. Each RIM also contains $\theta_{k}$ parameters that are shared between time steps. \\
However, only when the input is relevant are the RIM modules active and updated. For each time step, the attention selects and then activates a subset of the RIMs. Soft attention takes the product of a query represented as a matrix of dimension $N_{r} \times d$, with $d$ being the dimension of each key, and a collection of $N_{0}$ objects, each associated with a key as a row in matrix $K^{T} (N_{0} \times d$), and after normalizing (using Softmax) produces the following outputs: \\
\begin{equation}
\mathbf{Attention}(Q, K, V)=\mathbf{softmax}\left(\frac{Q K^{T}}{\sqrt{d}}\right) V,
\end{equation}
\\

where Q, K and V are the query, key and value matrices, respectively. The Softmax algorithm is then performed on each row of the argument matrix, resulting in a set of convex weights. As a consequence, a convex combination of the values in the rows of $V$ is obtained. Note that when the attention is focused on one element of a specific row, which means that the softmax is saturated, it selects one of the objects and copies its value to row j of the result. Further, the $d$ dimensions of the keys may be divided into heads, each of which has its own attention matrix and writes independently calculated values. \\

Without attention, neurons in neural networks operate on fixed variables and are thus fed by the previous layer. The key-value attention mechanism enables a dynamic selection of which a variable instance will be used as the input for each of the dynamics of the RIM’s arguments. These inputs may originate from an external source or be the output of another RIM. Thus, the model learns to dynamically choose those RIMs that are relevant to the present input. The RIMs provide the   queries for this specific application of key-value attention, while the current input provides the keys and values. The following summarizes the input attention paid to a particular RIM: \\

\begin{figure}[h]
\centering
\includegraphics[width=0.8\textwidth]{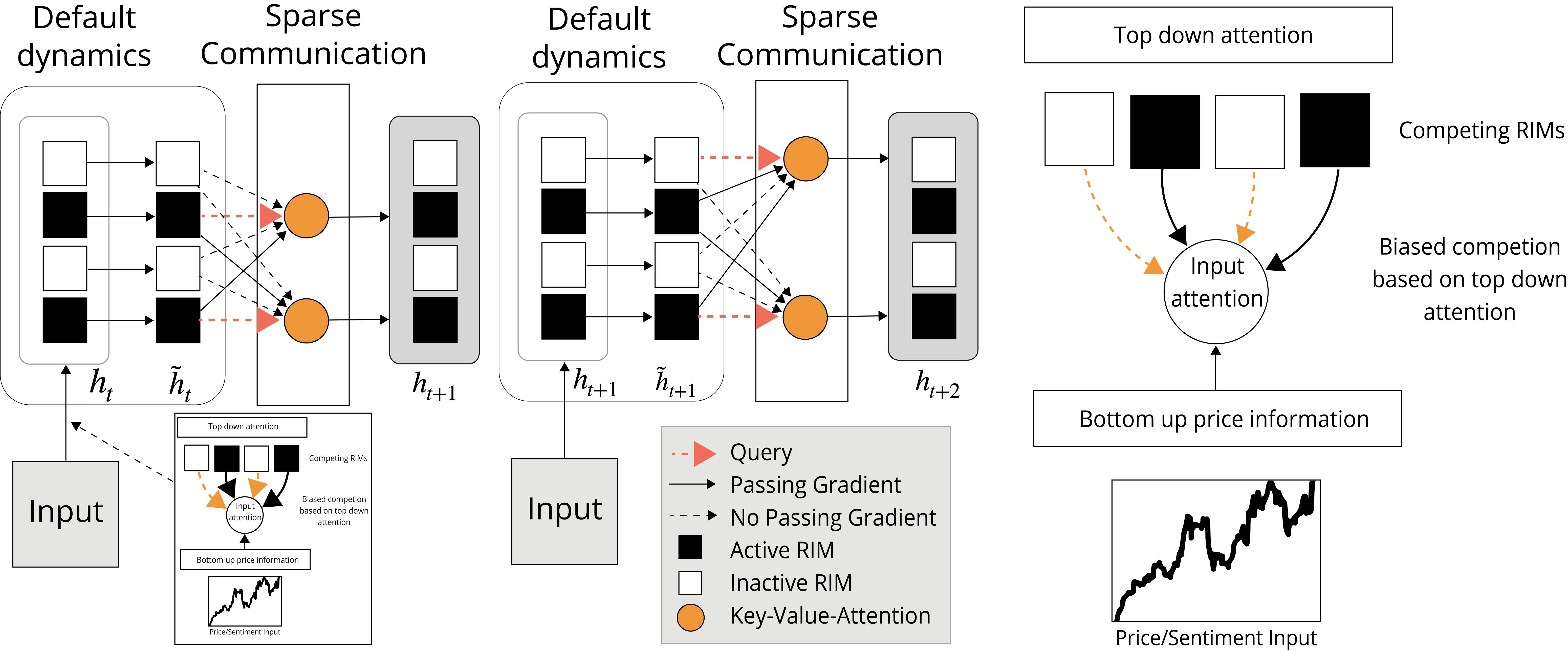}
\caption{Illustration of recurrent independent mechanisms.}
\label{fig:RIM}
\end{figure}

At time $t$, the input $x_t$ is seen as a collection of rows of a matrix, followed by the concatenation of a row of zeros to obtain the following equation:

\begin{equation}
X=\emptyset \oplus x_{t}
\end{equation} \\
Following this, linear transformations are applied to generate keys ($K=X W^{e}$, one for each input element and the null element), values ($V=X W^{v}$, one for each element), and queries ($Q=h_{t} W_{k}^{q}$, one for each RIM attention head). $W^v$ is a simple matrix translating an input element to the weighted attention's associated value vector, and $W^{e}$ is similarly a weight matrix mapping the input to the keys. $W_{k}^{q}$ is a weight matrix for each RIM that relates the RIM's hidden state to its queries. $\bigoplus$ refers to the concatenation operator at the row level. Thus, the attention can be written as follows:

\begin{equation}
A_{k}^{(i n)}=\mathbf{softmax}\left(\frac{h_{t} W_{k}^{q}\left(X W^{e}\right)^{T}}{\sqrt{d_{s}}}\right) X W^{v}, \mathbf{ where } \enspace \theta_{k}^{(i n)}=\left(W_{k}^{q}, W^{e}, W^{v}\right)
\end{equation}
\\
The RIMs use multiple heads for the input and communication attention, analogously to "Attention is all you need" \citep{DBLP:journals/corr/VaswaniSPUJGKP17}. In general, the RIMs operate independently by default and the attention mechanism allows the model to share information among the modules. Furthermore, the activated RIMs are allowed to read from all the other RIMs’ inputs. The reason for this is that non-activated RIMs do not need to change their value because they are not related to the current input. Nonetheless, they may retain important contextual information, therefore, there is communication between the RIMs through the usage of residual connections, as described in the paper "Relational recurrent neural networks" \citep{santoro2018relational}. Figure \ref{fig:RIM} illustrates the described RIM dynamics. The illustrations were adapted from the original paper and partially modified to fit the use case of this work. \\

However, the RIMs were originally used with gated recurrent units \citep{chung2014empirical} and LSTM \citep{LSTM}, which we decided against, because the $\alpha$-RNN and $\alpha_{t}$-RNN are being a much simpler architecture, yet perfectly suitable for forecasting stationary and non-stationary time series, respectively. 

\subsection{ \texorpdfstring{$\alpha$-RNNs and $\alpha_{t}$-RNNs} \label{}}

The $\alpha$-RNNs are a generic family of exponentially smoothed RNNs that excel in  modeling non-stationary time series data as seen in financial applications. They characterize the time series’ non-linear partial autocorrelation structure and directly capture dynamic influences such as seasonality and trends. The $\alpha$-RNN is almost identical to a standard RNN except for the addition of a scalar smoothing parameter, which provides the recurrent network with extended memory, that is, autoregressive memory beyond the sequence length.

To extend RNNs into dynamic time series models, the combination of the hidden state $\hat{h}_{t}$ and the exponentially smoothed output $\tilde{h}_{t}$, which is time-dependent and convex, is used. This combination means that the model is capable of modeling non-stationary time series data, as in the following equation: 
\begin{equation}
    \tilde{h}_{t+1}=\alpha_{t} \hat{h}_{t}+\left(1-\alpha_{t}\right) \tilde{h}_{t}
\end{equation}

Thus, smoothing may be thought of as a type of dynamic forecast error correction. Alternatively, smoothing may be
viewed as a weighted summation of the lagged data with either equal or lower weights, $\alpha_{t-s} \prod_{r=1}^{s}\left(1-\alpha_{t-r+1}\right)$ at the $s \geq 1$ lagged hidden state, $\hat{h}_{t-s}$:

\begin{equation}
    \tilde{h}_{t+1}=\alpha_{t} \hat{h}_{t}+\sum_{s=1}^{t-1} \alpha_{t-s} \prod_{r=1}^{s}\left(1-\alpha_{t-r+1}\right) \hat{h}_{t-s}+g(\alpha),
\end{equation}
where $g(\alpha):=\prod_{r=0}^{t-1}\left(1-\alpha_{t-r}\right) \tilde{y}_{1}$. Note that beyond the $r^{th}$ lag, for any $\alpha_{t-r+1}=1$, the model will forget the hidden states.

While the $\alpha_{t}$-RNN is free to specify how $\alpha$ is updated (including altering the update frequency) in response to the hidden state and input, using a recurrent layer is a convenient choice.

\subsection{ \texorpdfstring{$\alpha_{t}$-RIM} \label{}}

The activated RIMs in the $\alpha_{t}$-RIM uses the $\alpha_{t}$-RNN as their per-RIM independent transition dynamics. This choice was made because for industrial forecasting, LSTMs and GRUs are likely over-engineered, and light-weight exponentially smoothed architectures capture the key properties while being superior and more robust than simple RNNs.
\begin{figure}[h]
\centering
\includegraphics[width=0.8\textwidth]{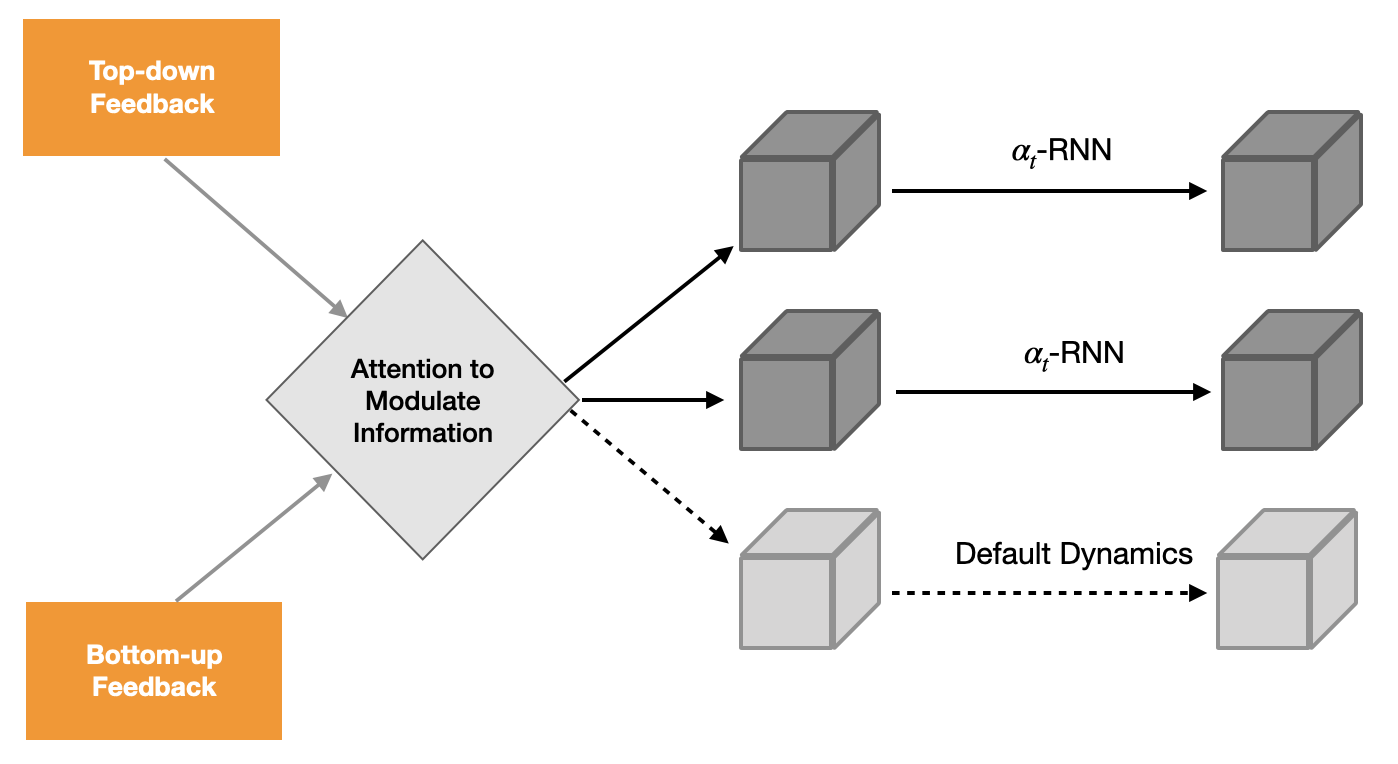}
\caption{Illustration of the $\alpha_{t}$-RIM functionality}
\label{fig:alpha_t-RIM}
\end{figure}

%\textbf{}: $\alpha$-RNNs are a generic family of exponentially smoothed RNNs, that excel in modeling non-stationary time series data as seen in financial applications. They characterize the time series' non-linear partial autocorrelation structure and directly capture dynamic influences such as seasonality and trends. The $\alpha$-RNN is nearly identical to a standard RNN except for the addition of a scalar smoothing parameter, which provides the recurrent network with extended memory. To extend the network to a dynamical time series model, it takes the combination, which is time-dependent and convex, of the hidden state $\hat{h}_{t}$ and of the exponentially smoothed output $\tilde{h}_{t}$ resulting in the following equation:

%\begin{equation}
%    \tilde{h}_{t+1}=\alpha_{t} \hat{h}_{t}+\left(1-\alpha_{t}\right) \tilde{h}_{t},
%\end{equation}

\section{Experiments and Results}\label{sec:others}
We evaluated our modeling approach using financial time series data with and without news sentiment, hereafter denoted as univariate and bivariate, respectively. To further study our model’s performance, we compared it with two RNN models, namely a simple RNN and an LSTM. TensorFlow v2.4.1 \citep{tensorflow2016} was used to implement all the models, while the $\alpha_{t}$-RIM was a custom implementation and the SimpleRNN and the LSTM from the TensorFlow-Keras-API were used.  Further, time series cross-validation  was performed using separate training, validation, and test sets for all the models and stocks. Each set represents a contiguous sample period, with the test set containing the most recent observations in order to maintain the data’s temporal structure and prevent look-ahead bias. The hidden layer was activated using Tanh functions. The Glorot and Bengio uniform \citep{pmlr-v9-glorot10a} approach was used to initialize the non-recurrent weight matrices, and an orthogonal matrix was used to initialize the recurrence weights for stability, which ensures that the absolute value of the eigenvalues is initially limited by unity  \citep{DBLP:journals/corr/HenaffSL16}. Further, Adam \citep{kingma2017adam} is used as optimizer. The hyperparameters for the RNN and LSTM were determined using time series cross-validation with five folds. The hidden units were evaluated starting from 5 up to and including 250 in steps of 5. L1 regularization was 0.0001, 0.001, 0.01, 0.10 and the dropout rate was 0.1, 0.2, 0.3, 0.4, 0.5, 0.6. The setup for the $\alpha_{t}$-RIM was identical, with the exception that the hyperparameters were determined with a three fold randomized grid search due to the 12 hyperparameters of the model. See Appendix Subsection 6.1 for more information on the model’s hyperparameters. Within that setup, we used three metrics: mean squared error (MSE), mean absolute error  (MAE), and mean percentage error (MAPE). The first two were used to evaluate the models during training, validation, and testing to examine their generalization performance on unseen data. The MAPE was used after training on the re-scaled testing dataset, where re-scaled means that the data transformations were reverted in order to retrieve the original closing prices, and separately for each prediction time step ahead, as our prediction horizon was five days ahead.

\subsection{Data}

Two data sources were used for the analyses: end of day pricing data from quandl.com and sentiment data from 2015 onwards, provided by YUKKA Lab, Berlin. The two datasets were then joined and contained 6.5 years of complete data from within the S\&P 500 Index. To account for variability in the experiments, trading volume, the total article count, and the number of positive and negative articles of all stocks were analyzed, resulting in the following chosen stocks:

\begin{enumerate}
\item \textbf {Amazon}: With high dollar volume and high article count. 
\item \textbf{Brown-Forman}: With low dollar volume and low article count. 
\item \textbf {Thermo Fisher}: With medium dollar volume and medium-high article count.
\end{enumerate}

To account for the noise in the sentiment data, we smoothed the sentiment score with a convolutional kernel filter \citep{convolutionFilter}, \citep{convSpline} during pre-processing. The datasets were then divided into training, validation, and testing sets. The training set was standardized using the mean and standard deviation of the training set and not the whole time series. Additionally, to avoid introducing systematic bias into the validation and test sets, identical normalization was used for the validation and test sets. In other words, the mean and standard deviation of the training set were used to normalize the validation and test sets. Further, log transformations were performed to decrease the data variability and bring the data closer to the normal distribution. Moreover, due to the strong downward impact of the COVID-19 pandemic on stock prices, the choice was made to remove the observation with the steep downward trend in the chosen stocks, as we expected the models to  perform poorly during this phase. This resulted in the following divisions: 

\begin{enumerate}
    \item \textbf {Training set}: 2015-01-05 – 2019-12-31
    \item \textbf{Validation set}: 2020-05-04 – 2020-12-28
    \item \textbf {Testing set}: 2020-12-29 – 2021-06-08 
\end{enumerate}

Moreover, we analyzed three different look-back time windows, 5, 10, and 21 days, to study the short- and long-term effects of the generalization related to the networks and time windows. 

\subsection{Results}
\begin{figure}[ht!]
  \centering
  \begin{minipage}[b]{0.48\textwidth}
    \includegraphics[width=\textwidth]{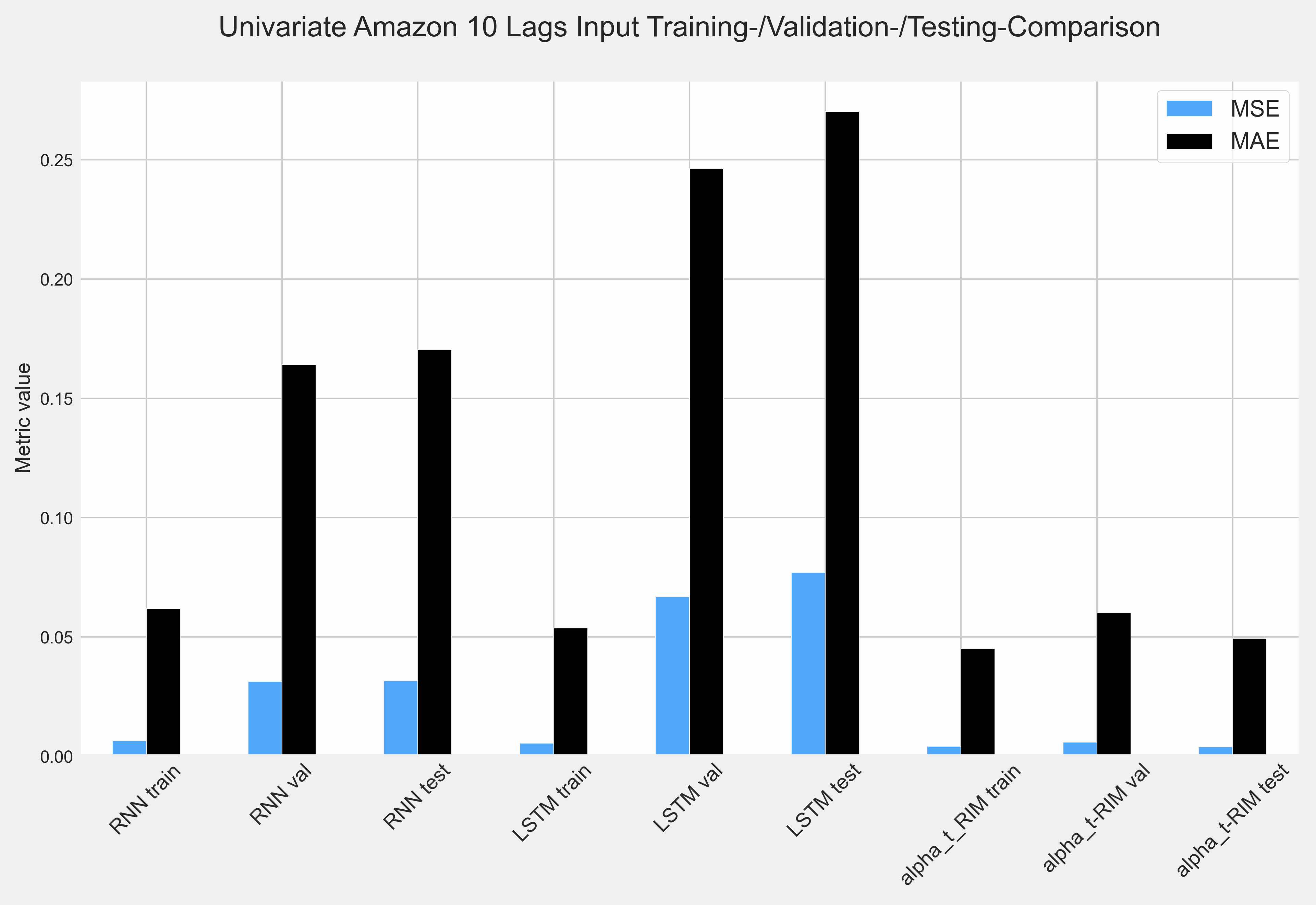}
    \caption{Univariate evaluation metrics of 10 lags input for Amazon stock.}\label{fig:uniAMZN10}
  \end{minipage}
  \hfill
  \begin{minipage}[b]{0.48\textwidth}
    \includegraphics[width=\textwidth]{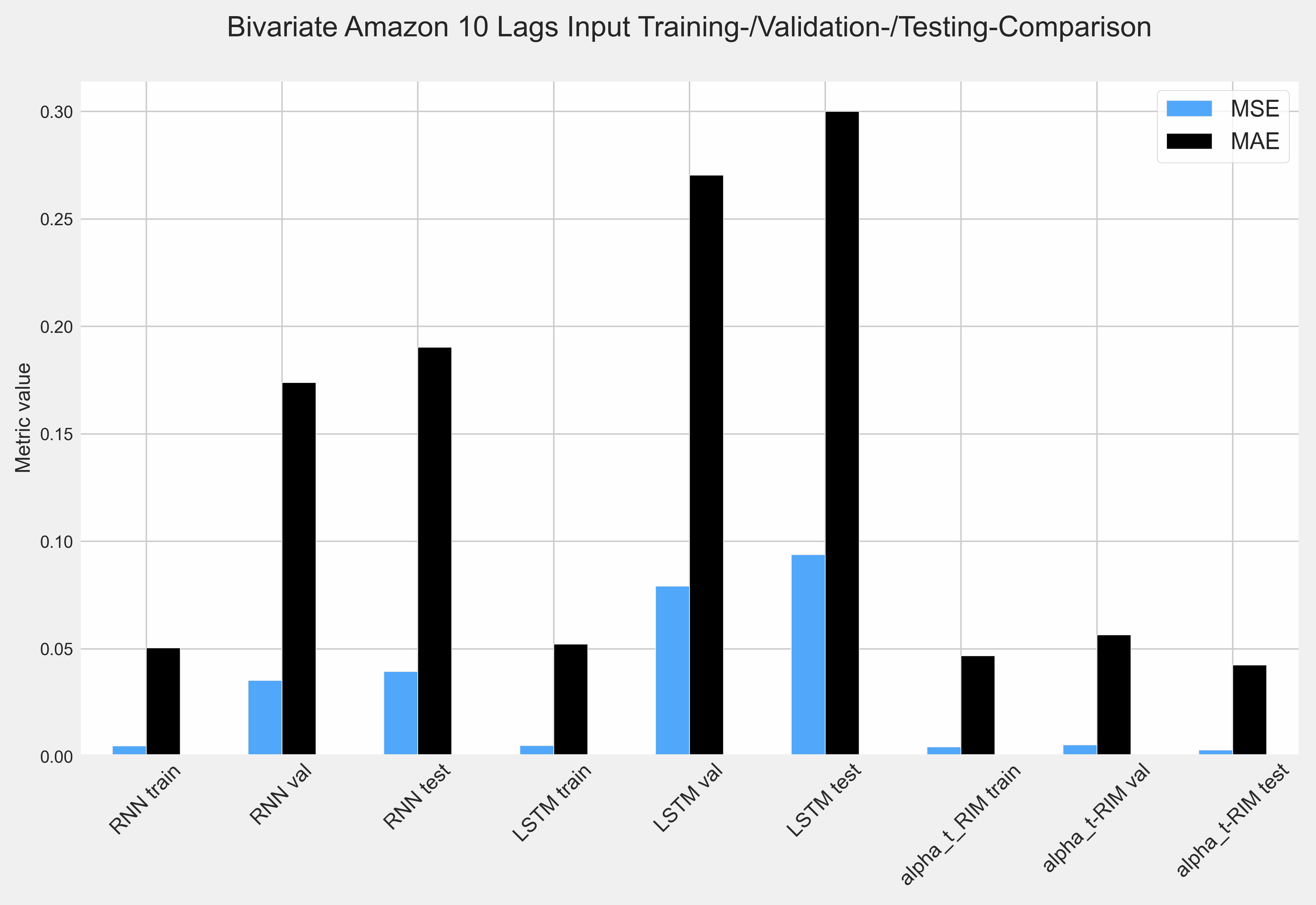}
    \caption{Bivariate evaluation metrics of 10 lags input for Amazon stock.}
  \end{minipage}
\end{figure}

\begin{figure}[ht!]
  \centering
  \begin{minipage}[b]{0.48\textwidth}
    \includegraphics[width=\textwidth]{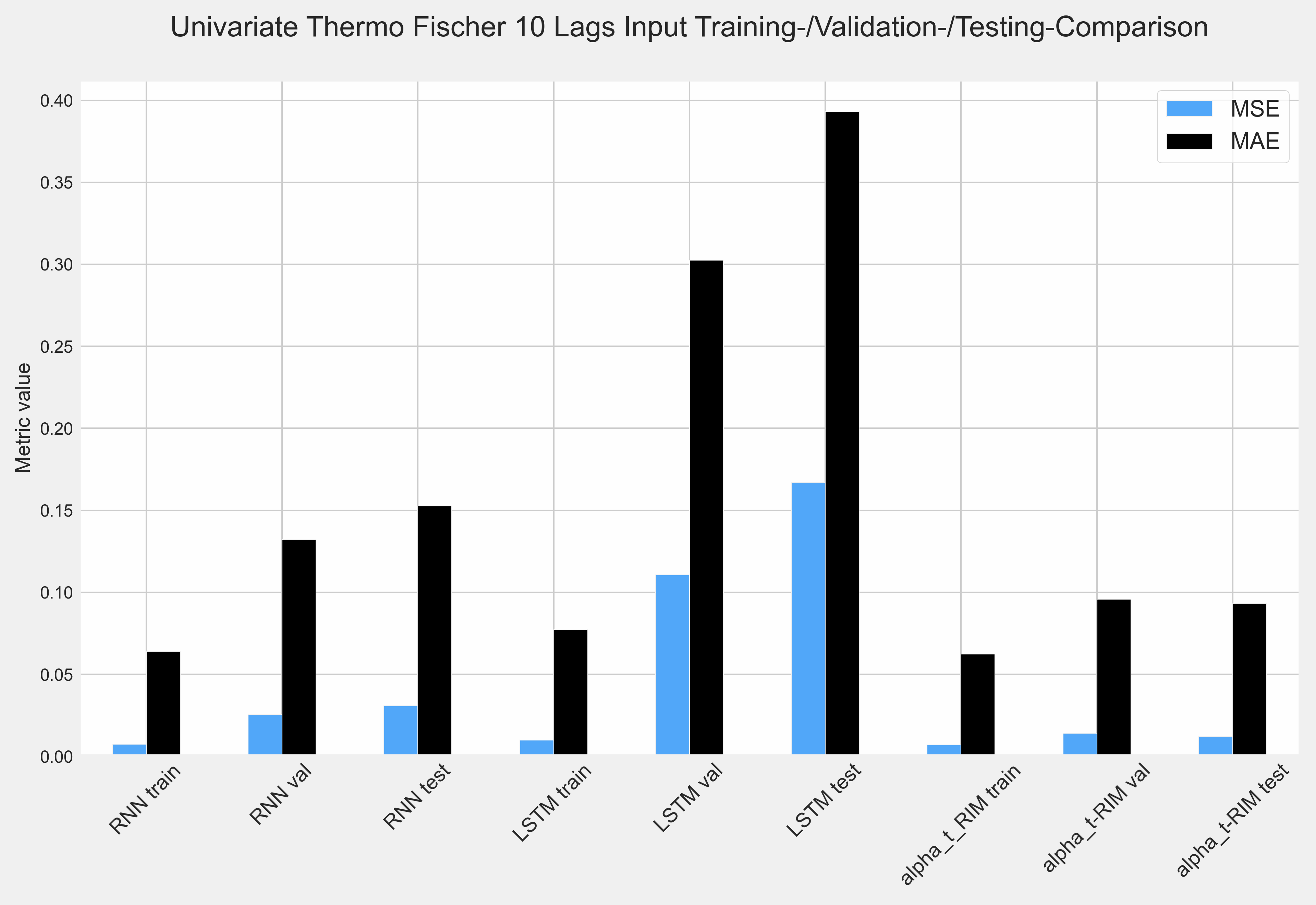}
    \caption{Univariate evaluation metrics of 10 lags input for Thermo Fischer stock.}
  \end{minipage}
  \hfill
  \begin{minipage}[b]{0.48\textwidth}
    \includegraphics[width=\textwidth]{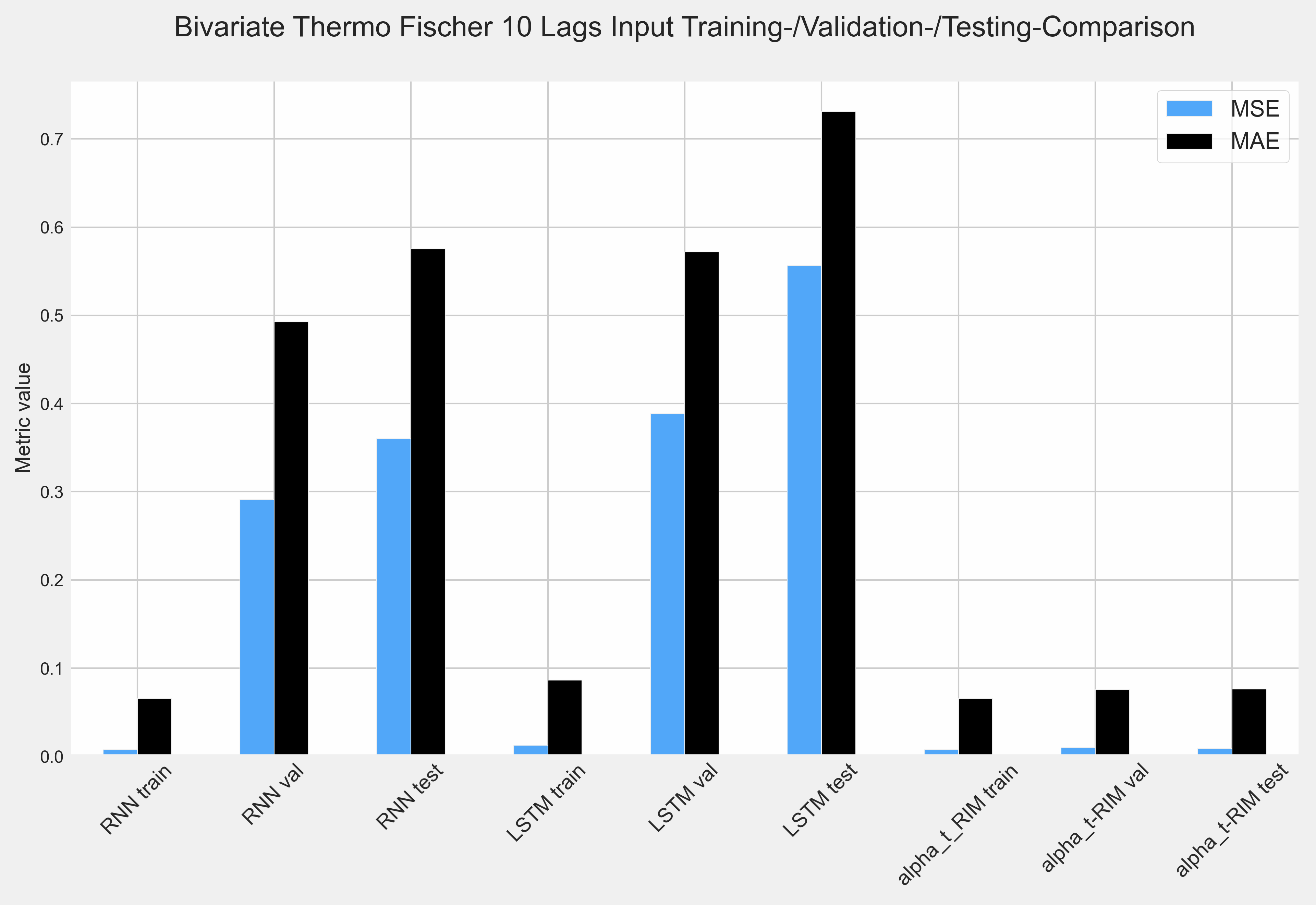}
    \caption{Bivariate evaluation metrics of 10 lags input for Thermo Fischer stock.}\label{fig:multiTMO10}
  \end{minipage}
\end{figure}

After comparing the results for all the networks, stocks, and time windows, we observed that the results depicted the same pattern. The performance of the training data sets differed little, that is, the MSE and MAE were almost identical during the training phase of the three different networks, namely the simple RNN, LSTM, and $\alpha_{t}$-RIM, but the values of the metrics increased steadily from the validation to the testing sets, for the simple RNN and LSTM. Accounting for the aforementioned results and the results from the comparison with each prediction step ahead led to the following two observations. Our model outperforms the simple RNN and the LSTM for each prediction time step and during the validation and testing phases of the networks. The Figures \ref{fig:uniAMZN10}-\ref{fig:multiTMO10} depict the results from the 10 day look-back and 5 day ahead prediction from the model evaluation and clearly demonstrate how the RNN and LSTM fail to generalize for the validation and testing sets. Furthermore, the $\alpha_{t}$-RIM is capable of using the sentiment score as an additional feature and improves its overall performance with the use of news sentiment. Regarding the input lags needed for the 5 days ahead prediction, the 10 lags resulted in the best overall performance for all tested networks, with two exceptions: the $\alpha_{t}$-RIM performance was slightly better with 21 lags input for Brown Forman stock, and the LSTM performed also better with 21 input lags, but for Thermo Fischer stock. A comparison of the results for Brown Forman stock with 10 input lags is presented in Tables \ref{tab:rescaled1} and \ref{tab:rescaled2} with the best values in bold.
\vspace{1.0 cm}

\vspace{1.0 cm}
\begin{figure}[ht!]
  \centering
  \begin{minipage}[b]{0.48\textwidth}
    \includegraphics[width=\textwidth]{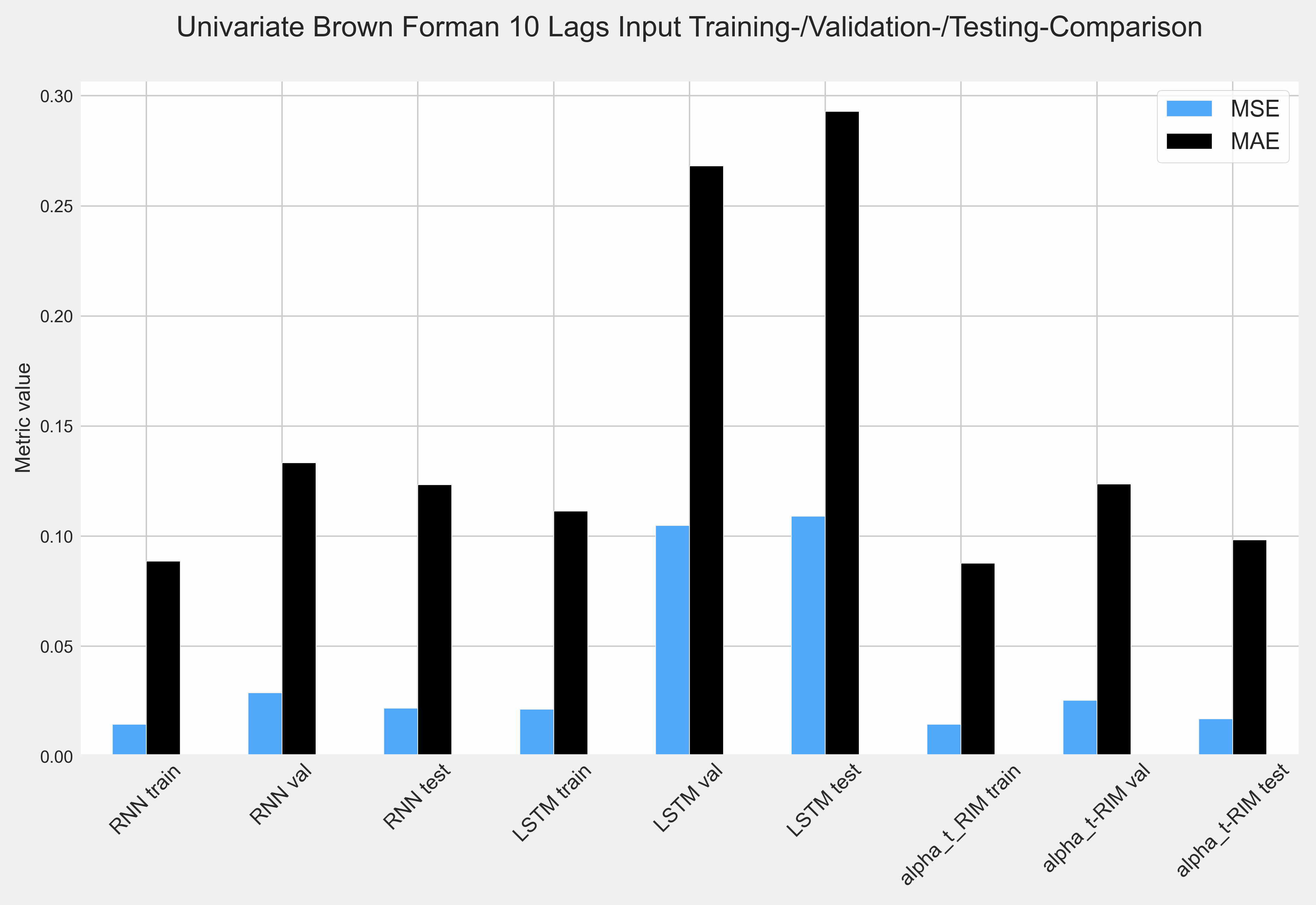}
    \caption{Univariate evaluation metrics of 10 lags input for Brown Forman stock.}
  \end{minipage}
  \hfill
  \begin{minipage}[b]{0.48\textwidth}
    \includegraphics[width=\textwidth]{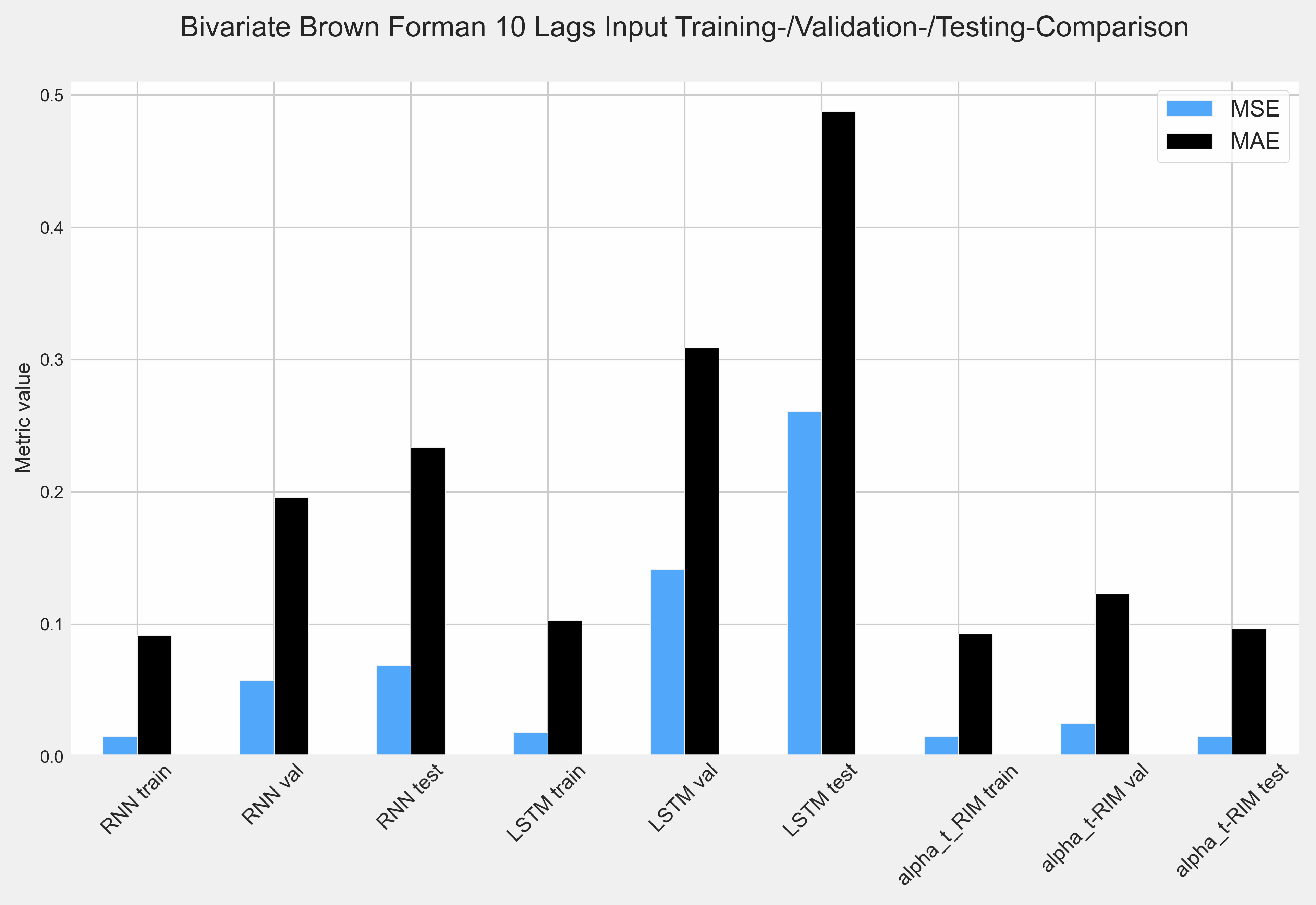}
    \caption{Bivariate evaluation metrics of 10 lags input for Brown Forman stock.}
  \end{minipage}
\end{figure}
\clearpage

\vspace{0.5 cm}

\begin{table}[!htb]
\renewcommand*{\arraystretch}{1.1}
    \begin{minipage}{.45\linewidth}
      \centering
        \begin{tabular}{r r r r r r r} \toprule
\multicolumn{1}{c}{} &
  \multicolumn{1}{c}{\textbf{RNN}} &
  \multicolumn{1}{c}{\textbf{LSTM}} &
  \multicolumn{1}{c}{${\alpha_{t}}$\textbf{-RIM}} & \\
\multicolumn{1}{c}{\textbf{Lag}} &
  \multicolumn{1}{c}{\textbf{MAPE}} &
  \multicolumn{1}{c}{\textbf{MAPE}} &
  \multicolumn{1}{c}{\textbf{MAPE}} & \\ \hline\hline
\textbf{1}        & 2.3279 & 5.2203 & \textbf{1.7081}  \\
\textbf{2}        & 1.9341 & 5.3022 & 1.7780    \\
\textbf{3}        & 2.6795 &  5.7144 & 1.9815  \\
\textbf{4}        & 2.3332 & 6.3815 & 2.1937  \\
\textbf{5}        & 3.0264 & 5.9082 & 2.3718  \\ 
\bottomrule
\end{tabular}
      \caption{Re-scaled univariate metrics of 10 lags input for the Brown Forman stock.}\label{tab:rescaled1}
    \end{minipage}%
    \hspace{1.6cm}
    \begin{minipage}{.45\linewidth}
      \centering
\begin{tabular}{r r r r r r r} \toprule
\multicolumn{1}{c}{} &
  \multicolumn{1}{c}{\textbf{RNN}} &
  \multicolumn{1}{c}{\textbf{LSTM}} &
  \multicolumn{1}{c}{${\alpha_{t}}$\textbf{-RIM}} & \\
\multicolumn{1}{c}{\textbf{Lag}} &
  \multicolumn{1}{c}{\textbf{MAPE}} &
  \multicolumn{1}{c}{\textbf{MAPE}} &
  \multicolumn{1}{c}{\textbf{MAPE}} & \\ \hline\hline
\textbf{1}        & 4.4243 & 8.7381 & \textbf{1.3157} \\
\textbf{2}        & 3.7217 & 9.1994 & 1.6702  \\
\textbf{3}        & 5.9833& 9.5561 &  1.9839  \\
\textbf{4}        & 3.9705 & 9.5775 &  2.2366  \\
\textbf{5}        & 4.8024 & 9.6431 &  2.4917  \\ 
\bottomrule
\end{tabular}
      \caption{Re-scaled bivariate metrics of 10 lags input for the Brown Forman stock.}\label{tab:rescaled2}
    \end{minipage} 
\end{table}

\section{Conclusion and outlook}
This work demonstrates the effectiveness of combining an RIM with an exponentially smoothed RNN for modeling non-stationary time series data. As our findings indicate, the $\alpha_{t}$-RIM outperforms the simple RNN and the LSTM, particularly within the validation and testing phases. This observation leads to the inference that the simple RNN and LSTM are not able to model changes in the unseen data. This demonstrates that the $\alpha_{t}$-RIM can generalize effectively on unseen data. Additionally, the bivariate time series prediction results suggest that the $\alpha_{t}$-RIM is capable of extracting patterns from news sentiment and using them as an additional input within its communication pattern to further stabilize the prediction accuracy. This result substantiates our earlier statement that it is necessary to utilize a new modeling approach that takes advantage of key-value attention, an exponentially smoothed RNN, and subsystems to model financial time series data in a dynamic and context-dependent way. Furthermore, in addition to the more accurate predictions that it provided, the advantage of the $\alpha_{t}$-RIM is that the attention weights and activation pattern of the modules can be visualized to further study the behavior of the model. In closing, for further research, we suggest investigating the possibility of including DSelect-k \citep{DkSelect} for the $\alpha_{t}$-RIM, which would the model allow to learn to select a different number of modules at each time-step. \\

\textbf{Acknowledgements:} The author would like to thank YUKKA Lab, Berlin, for providing the raw data for this research. Furthermore, the author would like to thank Matthew Dixon and Saeed Amen, who provided significant support to the research with their insights and expertise. Finally, the author would like to thank Jörg Osterrieder for his comments and suggestions on this paper.

\clearpage

%Bibliography
\bibliographystyle{unsrt}  
\bibliography{references}  

\clearpage

\section{Appendix}
\subsection{The \texorpdfstring{{${\alpha_{t}}$\textbf{-RIM}} Parameters}\label{hyperRIM}}

Due to the models’ constraints of the hyper-parameters (e.g., the number of RIMs have to be smaller or equal to $k$ modules), the normal cross-validation could not be performed. Therefore, a special function was implemented to generate a list of dictionaries to be fed into the grid search as a parameter grid. The list encompasses the following parameters:

\begin{itemize}
    \item \textbf{Units}: 2, 4, 6, 8, 10, 12, 14, 16, 18, 20, 25, 30, 35, 40, 45, 50
    \item \textbf{Number of RIMs}: 4, 6, 8, 10, 12, 14
    \item \textbf{K modules}:  4, 6, 8, 10, 12, 14 
    \item \textbf{Input key size}: 4, 6, 8, 10, 12 
    \item \textbf{Input value size}: 4, 6, 8, 10, 12
    \item \textbf{input query size}: 4, 6, 8, 10, 12
    \item \textbf{Input keep probability}: 0.6, 0.7, 0.8, 0.9
    \item \textbf{Number of communication heads}: 2, 4, 6, 8
    \item \textbf{Communication key size}: 4, 6, 8, 10, 12
    \item \textbf{Communication value size}: 4, 6, 8, 10, 12
    \item \textbf{Communication query size}: 4, 6, 8, 10, 12
    \item \textbf{Communication keep probability}: 0.6, 0.7, 0.8, 0.9
\end{itemize}

\subsection{Complete Training Results}
\vspace{0.5cm}

%=========================== Evaluation Metrics =================================
\subsubsection{Evaluation Metrics} 
\vspace{0.3cm}

%=========================== AMAZON =============================================
\textbf{AMAZON} \\
%############################# 5 input lags ######################################

\textbf{5 Input Lags} \\
\noindent
{\color{lightgray} \rule{\linewidth}{0.2mm} }

\begin{table}[!htb]
\renewcommand*{\arraystretch}{1.1}
    \begin{minipage}{.45\linewidth}
      \centering
        \begin{tabular}{lll}\toprule
                              \textbf{Model}            & \textbf{MSE} & \textbf{MAE} \\ \hline\hline
                            \textbf{RNN train}          & 0.0043       & 0.0476       \\
                            \textbf{RNN val}            & 0.0164       & 0.1123       \\
                \textbf{RNN test}                       & 0.0158       & 0.1158       \\
                \textbf{LSTM train}                     & 0.0049       & 0.0508       \\
                \textbf{LSTM val}                       & 0.0778       & 0.2683       \\
                \textbf{LSTM test}                      & 0.0916       & 0.2981       \\
\textbf{${\alpha_{t}}$\textbf{-RIM} train}   & \textbf{0.0045}       & \textbf{0.0484}       \\
\textbf{${\alpha_{t}}$\textbf{-RIM} val}     & \textbf{0.0048}       & \textbf{0.0536}       \\
\textbf{${\alpha_{t}}$\textbf{-RIM} test}    & \textbf{0.0028}       & \textbf{0.0416}       \\\bottomrule
\end{tabular}
      \caption{Univariate evaluation metrics of 5 lags input for the Amazon stock.}
    \end{minipage}%
    \hspace{1.6cm}
    \begin{minipage}{.45\linewidth}
      \centering
        \begin{tabular}{lll}\toprule
                \textbf{Model}     & \textbf{MSE} & \textbf{MAE} \\ \hline\hline
                                    \textbf{RNN train}   & 0.0065       & 0.0610       \\
                                    \textbf{RNN val}     & 0.0404       & 0.1894       \\
                                    \textbf{RNN test}    & 0.0433       & 0.2013       \\
                                    \textbf{LSTM train}  & 0.0049       & 0.0516       \\
                                    \textbf{LSTM val}    & 0.1044       & 0.3133       \\
                                    \textbf{LSTM test}   & 0.1259       & 0.3504       \\
\textbf{${\alpha_{t}}$\textbf{-RIM} train}    & \textbf{0.0055}       & {0.0535}       \\
\textbf{${\alpha_{t}}$\textbf{-RIM} val}      & \textbf{0.0050}        & \textbf{0.0547}       \\
\textbf{${\alpha_{t}}$\textbf{-RIM} test}     & \textbf{0.0027}       & \textbf{0.0406}       \\\bottomrule
\end{tabular}
      \caption{Bivariate evaluation metrics of 5 lags input for the Amazon stock.}
    \end{minipage} 
\end{table}

\clearpage
%############################# 10 input lags #####################################
\textbf{10 Input Lags}\\
\noindent
{\color{lightgray} \rule{\linewidth}{0.2mm} }
\begin{table}[!htb]
\renewcommand*{\arraystretch}{1.1}
    \begin{minipage}{.45\linewidth}
      \centering
        \begin{tabular}{lll}\toprule
              \textbf{Model}      & \textbf{MSE} & \textbf{MAE} \\ \hline\hline
                                    \textbf{RNN train}   & 0.0065       & 0.0620       \\
                                    \textbf{RNN val}     & 0.0314       & 0.1643       \\
                                    \textbf{RNN test}    & 0.0317       & 0.1704       \\
                                    \textbf{LSTM train}  & 0.0055       & 0.0538       \\
                                    \textbf{LSTM val}    & 0.0668       & 0.2463       \\
                                    \textbf{LSTM test}   & 0.077        & 0.2703       \\
\textbf{${\alpha_{t}}$\textbf{-RIM} train}    & \textbf{0.0042}       & \textbf{0.0451}       \\
\textbf{${\alpha_{t}}$\textbf{-RIM} val}      & \textbf{0.0059}       & \textbf{0.0601}       \\
\textbf{${\alpha_{t}}$\textbf{-RIM} test}     & \textbf{0.0039}       & \textbf{0.0495} \\\bottomrule
\end{tabular}
      \caption{Univariate evaluation metrics of 10 lags input the Amazon stock.}
    \end{minipage}%
    \hspace{1.6cm}
    \begin{minipage}{.45\linewidth}
      \centering
        \begin{tabular}{lll}\toprule
                \textbf{Model}     & \textbf{MSE} & \textbf{MAE} \\ \hline\hline
\textbf{RNN train}   & 0.0048       & 0.0504       \\
\textbf{RNN val}     & 0.0353       & 0.1738       \\
\textbf{RNN test}    & 0.0394       & 0.1903       \\
\textbf{LSTM train}  & 0.005        & 0.0522       \\
\textbf{LSTM val}    & 0.0791       & 0.2703       \\
\textbf{LSTM test}   & 0.0939       & 0.3       \\
\textbf{${\alpha_{t}}$\textbf{-RIM} train} & \textbf{0.0043}       & \textbf{0.0468}       \\
\textbf{${\alpha_{t}}$\textbf{-RIM} val}   & \textbf{0.0053}       & \textbf{0.0565}       \\
\textbf{${\alpha_{t}}$\textbf{-RIM} test}  & \textbf{0.0030}       & \textbf{0.0424} \\\bottomrule
\end{tabular}
      \caption{Bivariate evaluation metrics of 10 lags input for the Amazon stock.}
    \end{minipage} 
\end{table}

\vspace{1cm}
%############################# 21 input lags #####################################
\textbf{21 Input Lags}\\
\noindent
{\color{lightgray} \rule{\linewidth}{0.2mm} }
\begin{table}[!htb]
\renewcommand*{\arraystretch}{1.2}
    \begin{minipage}{.45\linewidth}
      \centering
        \begin{tabular}{lll}\toprule
                                      \textbf{Model}      & \textbf{MSE} & \textbf{MAE} \\ \hline\hline
                                    \textbf{RNN train}   & 0.0048       & 0.0503       \\
                                    \textbf{RNN val}     & 0.0148       & 0.1044       \\
                                    \textbf{RNN test}    & 0.0129       & 0.1006       \\
                                    \textbf{LSTM train}  & 0.0062       & 0.0590       \\
                                    \textbf{LSTM val}    & 0.0969       & 0.3011       \\
                                    \textbf{LSTM test}   & 0.1038       & 0.3141       \\
\textbf{${\alpha_{t}}$\textbf{-RIM} train}    & \textbf{0.0040}        & \textbf{0.0444}       \\
\textbf{${\alpha_{t}}$\textbf{-RIM} val}      & \textbf{0.0086}       & \textbf{0.0749}       \\
\textbf{${\alpha_{t}}$\textbf{-RIM} test}     & \textbf{0.0058}       & \textbf{0.0627}       \\\bottomrule
\end{tabular}
      \caption{Univariate evaluation metrics of 21 lags input for the Amazon stock.}
    \end{minipage}%
    \hspace{1.6cm}
    \begin{minipage}{.45\linewidth}
      \centering
        \begin{tabular}{lll}\toprule
                \textbf{Model}     & \textbf{MSE} & \textbf{MAE} \\ \hline\hline
                                        \textbf{RNN train}  & 0.0044       & 0.0474       \\
                                        \textbf{RNN val}     & 0.0125       & 0.0948       \\
                                        \textbf{RNN test}    & 0.0107       & 0.0917       \\
                                        \textbf{LSTM train}  & 0.0059       & 0.0590       \\
                                        \textbf{LSTM val}    & 0.1893       & 0.4250       \\
                                        \textbf{LSTM test}   & 0.2156       & 0.4562       \\
        \textbf{${\alpha_{t}}$\textbf{-RIM} train} & \textbf{0.0038}       & \textbf{0.0441}       \\
        \textbf{${\alpha_{t}}$\textbf{-RIM} val}   & \textbf{0.0061}       & \textbf{0.0612}       \\
        \textbf{${\alpha_{t}}$\textbf{-RIM} test}  & \textbf{0.0028}       & \textbf{0.0415 }      \\\bottomrule
\end{tabular}
      \caption{Bivariate evaluation metrics of 21 lags input for the Amazon stock.}
    \end{minipage} 
\end{table}

\clearpage
%=========================== BROWN FORMAN =======================================

\textbf{BROWN FORMAN} \\

%############################# 5 input lags ######################################

\textbf{5 Input Lags}\\
\noindent
{\color{lightgray} \rule{\linewidth}{0.2mm} }
\begin{table}[!htb]
\renewcommand*{\arraystretch}{1.0}
    \begin{minipage}{.45\linewidth}
      \centering
        \begin{tabular}{lll}\toprule
              \textbf{Model}      & \textbf{MSE} & \textbf{MAE} \\ \hline\hline
                                        \textbf{RNN train}   & 0.0156       & 0.091       \\
                                        \textbf{RNN val}     & 0.0345       & 0.1467       \\
                                        \textbf{RNN test}    & 0.0289       & 0.1394       \\
                                        \textbf{LSTM train}  & 0.0166       & 0.0950       \\
                                        \textbf{LSTM val}    & 0.0550        & 0.1897       \\
                                        \textbf{LSTM test}   & 0.0518       & 0.1975       \\
\textbf{${\alpha_{t}}$\textbf{-RIM} train}        & \textbf{0.014}        & \textbf{0.0865}       \\
\textbf{${\alpha_{t}}$\textbf{-RIM} val}          & \textbf{0.0225}       & \textbf{0.1146}       \\
\textbf{${\alpha_{t}}$\textbf{-RIM} test}         & \textbf{0.0164}       & \textbf{0.0967} \\\bottomrule
\end{tabular}
      \caption{Univariate evaluation metrics of 5 lags input for the Brown Forman stock.}
    \end{minipage}%
    \hspace{1.2cm}
    \begin{minipage}{.45\linewidth}
      \centering
        \begin{tabular}{lll}\toprule
                \textbf{Model}     & \textbf{MSE} & \textbf{MAE} \\ \hline\hline
                                        \textbf{RNN train}   & 0.0153       & 0.0914       \\
                                        \textbf{RNN val}     & 0.0392       & 0.1571       \\
                                        \textbf{RNN test}    & 0.0435       & 0.1804       \\
                                        \textbf{LSTM train}  & 0.0177       & 0.1011       \\
                                        \textbf{LSTM val}    & 0.1001       & 0.2575       \\
                                        \textbf{LSTM test}   & 0.1720        & 0.3885       \\
\textbf{${\alpha_{t}}$\textbf{-RIM} train}        & \textbf{0.0145}       & \textbf{0.0901}       \\
\textbf{${\alpha_{t}}$\textbf{-RIM} val}          & \textbf{0.0207}       & \textbf{0.1096}       \\
\textbf{${\alpha_{t}}$\textbf{-RIM} test}         & \textbf{0.0153}       & \textbf{0.1002} \\\bottomrule
\end{tabular}
      \caption{Bivariate evaluation metrics of 5 lags input for the Brown Forman stock.}
    \end{minipage} 
\end{table}

%############################# 10 input lags #####################################
\textbf{10 Input Lags}\\
\noindent
{\color{lightgray} \rule{\linewidth}{0.2mm} }
\begin{table}[!htb]
\renewcommand*{\arraystretch}{1.0}
    \begin{minipage}{.45\linewidth}
      \centering
        \begin{tabular}{lll}\toprule
              \textbf{Model}      & \textbf{MSE} & \textbf{MAE} \\ \hline\hline
                                        \textbf{RNN train}   & 0.0146       & 0.0887       \\
                                        \textbf{RNN val}     & 0.0289       & 0.1334       \\
                                        \textbf{RNN test}    & 0.0219       & 0.1234       \\
                                        \textbf{LSTM train}  & 0.0214       & 0.1114       \\
                                        \textbf{LSTM val}    & 0.1048       & 0.2682       \\
                                        \textbf{LSTM test}   & 0.1090        & 0.2929       \\
\textbf{${\alpha_{t}}$\textbf{-RIM} train}        & \textbf{0.0145}       & \textbf{0.0877}       \\
\textbf{${\alpha_{t}}$\textbf{-RIM} val}          & \textbf{0.0254}       & \textbf{0.1237}       \\
\textbf{${\alpha_{t}}$\textbf{-RIM} test}         & \textbf{0.0171}       & \textbf{0.0983} \\\bottomrule
\end{tabular}
      \caption{Univariate evaluation metrics of 10 lags input for the Brown Forman stock.}
    \end{minipage}%
    \hspace{1.2cm}
    \begin{minipage}{.45\linewidth}
      \centering
        \begin{tabular}{lll}\toprule
                \textbf{Model}     & \textbf{MSE} & \textbf{MAE} \\ \hline\hline
                                        \textbf{RNN train}   & 0.015        & 0.0912       \\
                                        \textbf{RNN val}     & 0.0570        & 0.1957       \\
                                        \textbf{RNN test}    & 0.0685       & 0.2333       \\
                                        \textbf{LSTM train}  & 0.0179       & 0.1026       \\
                                        \textbf{LSTM val}    & 0.1411       & 0.3087       \\
                                        \textbf{LSTM test}   & 0.2609       & 0.4876       \\
\textbf{${\alpha_{t}}$\textbf{-RIM} train}        & \textbf{0.0152}       & \textbf{0.0925}       \\
\textbf{${\alpha_{t}}$\textbf{-RIM} val}          & \textbf{0.0248}       & \textbf{0.1228}      \\
\textbf{${\alpha_{t}}$\textbf{-RIM} test}         & \textbf{0.0151}       & \textbf{0.0962} \\\bottomrule
\end{tabular}
      \caption{Bivariate evaluation metrics of 10 lags input for the Brown Forman stock.}
    \end{minipage} 
\end{table}

%############################# 21 input lags #####################################
\textbf{21 Input Lags}\\
\noindent
{\color{lightgray} \rule{\linewidth}{0.2mm} }
\begin{table}[!htb]
\renewcommand*{\arraystretch}{1.0}
    \begin{minipage}{.45\linewidth}
      \centering
        \begin{tabular}{lll}\toprule
              \textbf{Model}      & \textbf{MSE} & \textbf{MAE} \\ \hline\hline
                                    \textbf{RNN train}   & 0.0166       & 0.0946       \\
                                    \textbf{RNN val}     & 0.0561       & 0.1924       \\
                                    \textbf{RNN test}    & 0.0556       & 0.2054       \\
                                    \textbf{LSTM train}  & 0.0247       & 0.1210       \\
                                    \textbf{LSTM val}    & 0.1739       & 0.3700       \\
                                    \textbf{LSTM test}   & 0.1872       & 0.4011       \\
\textbf{${\alpha_{t}}$\textbf{-RIM} train}    & \textbf{0.0136}       & \textbf{0.0843}       \\
\textbf{${\alpha_{t}}$\textbf{-RIM} val}      & \textbf{0.0208}       & \textbf{0.1091}       \\
\textbf{${\alpha_{t}}$\textbf{-RIM} test}     & \textbf{0.0130}        & \textbf{0.0884}       \\\bottomrule
\end{tabular}
      \caption{Univariate evaluation metrics of 21 lags input for the Brown Forman stock.}
    \end{minipage}%
    \hspace{1.2cm}
    \begin{minipage}{.45\linewidth}
      \centering
        \begin{tabular}{lll}\toprule
                \textbf{Model}     & \textbf{MSE} & \textbf{MAE} \\ \hline\hline
                                        \textbf{RNN train}   & 0.0161       & 0.0942       \\
                                        \textbf{RNN val}     & 0.0707       & 0.2219       \\
                                        \textbf{RNN test}    & 0.0846       & 0.2624       \\
                                        \textbf{LSTM train}  & 0.0254       & 0.1253       \\
                                        \textbf{LSTM val}    & 0.2590        & 0.4507       \\
                                        \textbf{LSTM test}   & 0.3775       & 0.5908       \\
\textbf{${\alpha_{t}}$\textbf{-RIM} train}        & \textbf{0.0136}       & \textbf{0.0854}       \\
\textbf{${\alpha_{t}}$\textbf{-RIM} val}          & \textbf{0.0237}       & \textbf{0.1194}       \\
\textbf{${\alpha_{t}}$\textbf{-RIM} test}         & \textbf{0.0187}       & \textbf{0.1159} \\\bottomrule
\end{tabular}
      \caption{Bivariate evaluation metrics of 21 lags input for the Brown Forman stock.}
    \end{minipage} 
\end{table}

\clearpage
%=========================== BROWN FORMAN =======================================

\textbf{THERMO FISCHER} \\

%############################# 5 input lags ######################################

\textbf{5 Input Lags}\\
\noindent
{\color{lightgray} \rule{\linewidth}{0.2mm} }
\begin{table}[!htb]
\renewcommand*{\arraystretch}{1.0}
    \begin{minipage}{.45\linewidth}
      \centering
        \begin{tabular}{lll}\toprule
              \textbf{Model}      & \textbf{MSE} & \textbf{MAE} \\ \hline\hline
                                        \textbf{RNN train}   & 0.0076       & 0.0669       \\
                                        \textbf{RNN val}     & 0.066        & 0.2272       \\
                                        \textbf{RNN test}    & 0.1031       & 0.3013       \\
                                        \textbf{LSTM train}  & 0.0089       & 0.0737       \\
                                        \textbf{LSTM val}    & 0.1200         & 0.3146       \\
                                        \textbf{LSTM test}   & 0.2002       & 0.4345       \\
\textbf{${\alpha_{t}}$\textbf{-RIM} train}        & \textbf{0.0086}       & \textbf{0.0727}       \\
\textbf{${\alpha_{t}}$\textbf{-RIM} val}          & \textbf{0.0106}       & \textbf{0.0797}       \\
\textbf{${\alpha_{t}}$\textbf{-RIM} test}         & \textbf{0.0138}       & \textbf{0.0895} \\\bottomrule
\end{tabular}
      \caption{Univariate evaluation metrics of 5 lags input for the Thermo Fischer stock.}
    \end{minipage}%
    \hspace{1.2cm}
    \begin{minipage}{.45\linewidth}
      \centering
        \begin{tabular}{lll}\toprule
                \textbf{Model}     & \textbf{MSE} & \textbf{MAE} \\ \hline\hline
                                        \textbf{RNN train}   & 0.0073       & 0.0640       \\
                                        \textbf{RNN val}     & 0.2113       & 0.4069       \\
                                        \textbf{RNN test}    & 0.2637       & 0.4837       \\
                                        \textbf{LSTM train}  & 0.0091       & 0.0746       \\
                                        \textbf{LSTM val}    & 0.4518       & 0.6262       \\
                                        \textbf{LSTM test}   & 0.6633       & 0.8007       \\
\textbf{${\alpha_{t}}$\textbf{-RIM} train}        & \textbf{0.0079}       & \textbf{0.0685}       \\
\textbf{${\alpha_{t}}$\textbf{-RIM} val}          & \textbf{0.0125}       & \textbf{0.0889}       \\
\textbf{${\alpha_{t}}$\textbf{-RIM} test}         & \textbf{0.0108}       & \textbf{0.0803} \\\bottomrule
\end{tabular}
      \caption{Bivariate evaluation metrics of 5 lags input for the Thermo Fischer stock.}
    \end{minipage} 
\end{table}

%############################# 10 input lags #####################################
\textbf{10 Input Lags}\\
\noindent
{\color{lightgray} \rule{\linewidth}{0.2mm} }
\begin{table}[!htb]
\renewcommand*{\arraystretch}{1.0}
    \begin{minipage}{.45\linewidth}
      \centering
        \begin{tabular}{lll}\toprule
              \textbf{Model}      & \textbf{MSE} & \textbf{MAE} \\ \hline\hline
                                        \textbf{RNN train}   & 0.0073       & 0.0638       \\
                                        \textbf{RNN val}     & 0.0255       & 0.1321       \\
                                        \textbf{RNN test}    & 0.0307       & 0.1526       \\
                                        \textbf{LSTM train}  & 0.0098       & 0.0775       \\
                                        \textbf{LSTM val}    & 0.1107       & 0.3025       \\
                                        \textbf{LSTM test}   & 0.1670        & 0.3932       \\
\textbf{${\alpha_{t}}$\textbf{-RIM} train}        & \textbf{0.0073}       & \textbf{0.0648}       \\
\textbf{${\alpha_{t}}$\textbf{-RIM} val}          & \textbf{0.0608}       & \textbf{0.2194}       \\
\textbf{${\alpha_{t}}$\textbf{-RIM} test}         & \textbf{0.0894}      & \textbf{0.2830} \\\bottomrule
\end{tabular}
      \caption{Univariate evaluation metrics of 10 lags input for the Thermo Fischer stock.}
    \end{minipage}%
    \hspace{1.2cm}
    \begin{minipage}{.45\linewidth}
      \centering
        \begin{tabular}{lll}\toprule
               \textbf{Model}     & \textbf{MSE} & \textbf{MAE} \\ \hline\hline
                                    \textbf{RNN train}   & 0.0075       & 0.0656       \\
                                    \textbf{RNN val}     & 0.2912       & 0.4925       \\
                                    \textbf{RNN test}    & 0.3599       & 0.5755       \\
                                    \textbf{LSTM train}  & 0.0126       & 0.0866       \\
                                    \textbf{LSTM val}    & 0.3885       & 0.5720       \\
                                    \textbf{LSTM test}   & 0.5568       & 0.7312       \\
\textbf{${\alpha_{t}}$\textbf{-RIM} train}    & \textbf{0.0077}       & \textbf{0.0656}       \\
\textbf{${\alpha_{t}}$\textbf{-RIM} val}      & \textbf{0.0097}       & \textbf{0.0757}       \\
\textbf{${\alpha_{t}}$\textbf{-RIM} test}     & \textbf{0.0090 }      & \textbf{0.0764} \\\bottomrule
\end{tabular}
      \caption{Bivariate evaluation metrics of 10 lags input for the Thermo Fischer stock.}
    \end{minipage} 
\end{table}

%############################# 21 input lags #####################################
\textbf{21 Input Lags}\\
\noindent
{\color{lightgray} \rule{\linewidth}{0.2mm} }
\begin{table}[!htb]
\renewcommand*{\arraystretch}{1.0}
    \begin{minipage}{.45\linewidth}
      \centering
        \begin{tabular}{lll}\toprule
              \textbf{Model}      & \textbf{MSE} & \textbf{MAE} \\ \hline\hline
                                        \textbf{RNN train}   & 0.0076       & 0.0653       \\
                                        \textbf{RNN val}     & 0.0461       & 0.1860       \\
                                        \textbf{RNN test}    & 0.0563       & 0.2120       \\
                                        \textbf{LSTM train}  & 0.0118       & 0.0818       \\
                                        \textbf{LSTM val}    & 0.0901       & 0.2643       \\
                                        \textbf{LSTM test}   & 0.1290        & 0.3421       \\
\textbf{${\alpha_{t}}$\textbf{-RIM} train}        & \textbf{0.0079}       & \textbf{0.0689}       \\
\textbf{${\alpha_{t}}$\textbf{-RIM} val}          & \textbf{0.0195}       & \textbf{0.1179}       \\
\textbf{${\alpha_{t}}$\textbf{-RIM} test}         & \textbf{0.0131}       & \textbf{0.0944} \\\bottomrule
\end{tabular}
      \caption{Univariate evaluation metrics of 21 lags input for the Thermo Fischer stock.}
    \end{minipage}%
    \hspace{1.2cm}
    \begin{minipage}{.45\linewidth}
      \centering
        \begin{tabular}{lll}\toprule
                \textbf{Model}     & \textbf{MSE} & \textbf{MAE} \\ \hline\hline
                                        \textbf{RNN train}   & 0.0074       & 0.0647       \\
                                        \textbf{RNN val}     & 0.192        & 0.3984       \\
                                        \textbf{RNN test}    & 0.2637       & 0.4837       \\
                                        \textbf{LSTM train}  & 0.1816       & 0.4037       \\
                                        \textbf{LSTM val}    & 0.4833       & 0.6598       \\
                                        \textbf{LSTM test}   & 0.5983       & 0.7646       \\
\textbf{${\alpha_{t}}$\textbf{-RIM} train}        & \textbf{0.0065}       & \textbf{0.0608}       \\
\textbf{${\alpha_{t}}$\textbf{-RIM} val}          & \textbf{0.0146}       & \textbf{0.0884}       \\
\textbf{${\alpha_{t}}$\textbf{-RIM} test}         & \textbf{0.0176}       & \textbf{0.1028} \\\bottomrule
\end{tabular}
      \caption{Bivariate evaluation metrics of 21 lags input for the Thermo Fischer stock.}
    \end{minipage}
\end{table}

%=========================== Re-scaled Metrics ================================

\subsubsection{Re-Scaled Metrics} 
\vspace{0.3cm}

\textbf{AMAZON} \\

%############################# 5 input lags Amazon #####################################
\textbf{5 Input Lags}\\
\noindent
{\color{lightgray} \rule{\linewidth}{0.2mm} }

\begin{table}[!htb]
\renewcommand*{\arraystretch}{1.2}
    \begin{minipage}{.45\linewidth}
      \centering
        \begin{tabular}{r r r r r r r} \toprule
\multicolumn{1}{c}{} &
  \multicolumn{1}{c}{\textbf{RNN}} &
  \multicolumn{1}{c}{\textbf{LSTM}} &
  \multicolumn{1}{c}{${\alpha_{t}}$\textbf{-RIM}} & \\
\multicolumn{1}{c}{\textbf{Lag}} &
  \multicolumn{1}{c}{\textbf{MAPE}} &
  \multicolumn{1}{c}{\textbf{MAPE}} &
  \multicolumn{1}{c}{\textbf{MAPE}} & \\ \hline\hline
\textbf{1}        & 6.578        & 14.9147       & \textbf{1.7589}                 \\
\textbf{2}        & 5.0371       & 14.3504       & 1.9456                 \\
\textbf{3}        & 6.7379       & 14.7175       & 2.1742                 \\
\textbf{4}        & 5.9735       & 14.0910        & 2.4410                  \\
\textbf{5}        & 5.2029       & 14.646        & 2.6204  \\ 
\bottomrule
\end{tabular}
      \caption{Re-scaled univariate metrics of 5 input lags for the Amazon stock.}
    \end{minipage}%
    \hspace{1.6cm}
    \begin{minipage}{.45\linewidth}
      \centering
\begin{tabular}{r r r r r r r} \toprule
\multicolumn{1}{c}{} &
  \multicolumn{1}{c}{\textbf{RNN}} &
  \multicolumn{1}{c}{\textbf{LSTM}} &
  \multicolumn{1}{c}{${\alpha_{t}}$\textbf{-RIM}} & \\
\multicolumn{1}{c}{\textbf{Lag}} &
  \multicolumn{1}{c}{\textbf{MAPE}} &
  \multicolumn{1}{c}{\textbf{MAPE}} &
  \multicolumn{1}{c}{\textbf{MAPE}} & \\ \hline\hline
\textbf{1}        & 9.8698       & 15.6884       & \textbf{1.6553}                 \\
\textbf{2}        & 8.5604       & 16.8496       & 1.9656                 \\
\textbf{3}        & 11.62        & 18.0037       & 2.1480                  \\
\textbf{4}        & 10.6241      & 16.8135       & 2.3762                 \\
\textbf{5}        & 9.6089       & 17.0085       & 2.5653  \\ 
\bottomrule
\end{tabular}
      \caption{Re-scaled bivariate metrics of 5 input lags for the Amazon stock.}
    \end{minipage} 
\end{table}

\vspace{0.3cm}
%############################# 10 input lags Amazon #####################################
\textbf{10 Input Lags}\\
\noindent
{\color{lightgray} \rule{\linewidth}{0.2mm} }

\begin{table}[!htb]
\renewcommand*{\arraystretch}{1.2}
    \begin{minipage}{.45\linewidth}
      \centering
        \begin{tabular}{r r r r r r r} \toprule
\multicolumn{1}{c}{} &
  \multicolumn{1}{c}{\textbf{RNN}} &
  \multicolumn{1}{c}{\textbf{LSTM}} &
  \multicolumn{1}{c}{${\alpha_{t}}$\textbf{-RIM}} & \\
\multicolumn{1}{c}{\textbf{Lag}} &
  \multicolumn{1}{c}{\textbf{MAPE}} &
  \multicolumn{1}{c}{\textbf{MAPE}} &
  \multicolumn{1}{c}{\textbf{MAPE}} & \\ \hline\hline
\textbf{1}        & 9.1246       & 11.4027       & \textbf{1.6793}                 \\
\textbf{2}        & 8.4374       & 13.8804       & 2.0933                 \\
\textbf{3}        & 9.0449       & 13.4336       & 2.5583                 \\
\textbf{4}        & 8.3960        & 15.2850        & 3.1922                 \\
\textbf{5}        & 7.8868       & 12.3275       & 3.3351  \\ 
\bottomrule
\end{tabular}
      \caption{Re-scaled univariate metrics of 10 input lags for the Amazon stock.}
    \end{minipage}%
    \hspace{1.6cm}
    \begin{minipage}{.45\linewidth}
      \centering
\begin{tabular}{r r r r r r r} \toprule
\multicolumn{1}{c}{} &
  \multicolumn{1}{c}{\textbf{RNN}} &
  \multicolumn{1}{c}{\textbf{LSTM}} &
  \multicolumn{1}{c}{${\alpha_{t}}$\textbf{-RIM}} & \\
\multicolumn{1}{c}{\textbf{Lag}} &
  \multicolumn{1}{c}{\textbf{MAPE}} &
  \multicolumn{1}{c}{\textbf{MAPE}} &
  \multicolumn{1}{c}{\textbf{MAPE}} & \\ \hline\hline
\textbf{1}        & 8.0755       & 12.3300         & \textbf{1.5730}                  \\
\textbf{2}        & 8.9933       & 16.0122       & 1.9777                 \\
\textbf{3}        & 11.4880       & 15.8318       & 2.2137                 \\
\textbf{4}        & 8.4545       & 15.0134       & 2.6374                 \\
\textbf{5}        & 10.6264      & 13.8912       & 2.7156  \\ 
\bottomrule
\end{tabular}
      \caption{Re-scaled bivariate metrics of 10 input lags for the Amazon stock.}
    \end{minipage} 
\end{table}

\vspace{0.3cm}

%############################# 21 input lags Amazon #####################################
\textbf{21 Input Lags}\\
\noindent
{\color{lightgray} \rule{\linewidth}{0.2mm} }

\begin{table}[!htb]
\renewcommand*{\arraystretch}{1.2}
    \begin{minipage}{.45\linewidth}
      \centering
        \begin{tabular}{r r r r r r r} \toprule
\multicolumn{1}{c}{} &
  \multicolumn{1}{c}{\textbf{rnn}} &
  \multicolumn{1}{c}{\textbf{lstm}} &
  \multicolumn{1}{c}{${\alpha_{t}}$\textbf{-RIM}} & \\
\multicolumn{1}{c}{\textbf{Lag}} &
  \multicolumn{1}{c}{\textbf{MAPE}} &
  \multicolumn{1}{c}{\textbf{MAPE}} &
  \multicolumn{1}{c}{\textbf{MAPE}} & \\ \hline\hline
\textbf{1}        & 6.6462       & 12.1883       & \textbf{2.0397}                 \\
\textbf{2}        & 3.9523       & 16.8744       & 2.8626                 \\
\textbf{3}        & 5.4141       & 17.4566       & 2.9411                 \\
\textbf{4}        & 4.7591       & 15.8323       & 3.8771                 \\
\textbf{5}        & 4.9461       & 13.8310        & 4.4702  \\  
\bottomrule
\end{tabular}
      \caption{Re-scaled univariate metrics of 21 input lags for the Amazon stock.}
    \end{minipage}%
    \hspace{1.6cm}
    \begin{minipage}{.45\linewidth}
      \centering
\begin{tabular}{r r r r r r r} \toprule
\multicolumn{1}{c}{} &
  \multicolumn{1}{c}{\textbf{RNN}} &
  \multicolumn{1}{c}{\textbf{LSTM}} &
  \multicolumn{1}{c}{${\alpha_{t}}$\textbf{-RIM}} & \\
\multicolumn{1}{c}{\textbf{Lag}} &
  \multicolumn{1}{c}{\textbf{MAPE}} &
  \multicolumn{1}{c}{\textbf{MAPE}} &
  \multicolumn{1}{c}{\textbf{MAPE}} & \\ \hline\hline
\textbf{1}        & 4.4279       & 16.6017       & \textbf{1.4672}                 \\
\textbf{2}        & 3.7260        & 22.7144       & 1.9301                 \\
\textbf{3}        & 6.3038       & 24.3148       & 2.1216                 \\
\textbf{4}        & 4.2521       & 21.8113       & 2.4899                 \\
\textbf{5}        & 4.7856       & 21.2809       & 2.8684  \\ 
\bottomrule
\end{tabular}
      \caption{Re-scaled bivariate metrics of 21 input lags for the Amazon stock.}
    \end{minipage} 
\end{table}

%=========================== BROWN FORMAN ======================================
\clearpage

\textbf{BROWN FORMAN} \\

%############################# 5 input lags Brown Forman #####################################
\textbf{5 Input Lags}\\
\noindent
{\color{lightgray} \rule{\linewidth}{0.2mm} }

\begin{table}[!htb]
\renewcommand*{\arraystretch}{1.2}
    \begin{minipage}{.45\linewidth}
      \centering
        \begin{tabular}{r r r r r r r} \toprule
\multicolumn{1}{c}{} &
  \multicolumn{1}{c}{\textbf{RNN}} &
  \multicolumn{1}{c}{\textbf{LSTM}} &
  \multicolumn{1}{c}{${\alpha_{t}}$\textbf{-RIM}} & \\
\multicolumn{1}{c}{\textbf{Lag}} &
  \multicolumn{1}{c}{\textbf{MAPE}} &
  \multicolumn{1}{c}{\textbf{MAPE}} &
  \multicolumn{1}{c}{\textbf{MAPE}} & \\ \hline\hline
\textbf{1}        & 4.5914       & 3.5285        & \textbf{1.4819}                 \\
\textbf{2}        & 1.8316       & 3.6831        & 1.7094                 \\
\textbf{3}        & 2.0918       & 3.9442        & 2.0129                 \\
\textbf{4}        & 2.1048       & 3.7976        & 2.2163                 \\
\textbf{5}        & 3.2806       & 4.5023        & 2.4563  \\ 
\bottomrule
\end{tabular}
      \caption{Re-scaled univariate metrics of 5 input lags for the Brown Forman stock.}
    \end{minipage}%
    \hspace{1.6cm}
    \begin{minipage}{.45\linewidth}
      \centering
\begin{tabular}{r r r r r r r} \toprule
\multicolumn{1}{c}{} &
  \multicolumn{1}{c}{\textbf{RNN}} &
  \multicolumn{1}{c}{\textbf{LSTM}} &
  \multicolumn{1}{c}{${\alpha_{t}}$\textbf{-RIM}} & \\
\multicolumn{1}{c}{\textbf{Lag}} &
  \multicolumn{1}{c}{\textbf{MAPE}} &
  \multicolumn{1}{c}{\textbf{MAPE}} &
  \multicolumn{1}{c}{\textbf{MAPE}} & \\ \hline\hline
\textbf{1}        & 2.7613       & 6.5932        & \textbf{1.3910}                  \\
\textbf{2}        & 3.0322       & 7.6978        & 1.7784                 \\
\textbf{3}        & 4.0114       & 8.1611        & 1.9711                 \\
\textbf{4}        & 3.6145       & 7.6531        & 2.2752                 \\
\textbf{5}        & 4.3920        & 7.4530         & 2.6458  \\ 
\bottomrule
\end{tabular}
      \caption{Re-scaled bivariate metrics of 5 input lags for the Brown Forman stock.}
    \end{minipage} 
\end{table}

\vspace{0.3cm}
%############################# 10 input lags Brown Forman #####################################
\textbf{10 Input Lags}\\
\noindent
{\color{lightgray} \rule{\linewidth}{0.2mm} }

\begin{table}[!htb]
\renewcommand*{\arraystretch}{1.2}
    \begin{minipage}{.45\linewidth}
      \centering
        \begin{tabular}{r r r r r r r} \toprule
\multicolumn{1}{c}{} &
  \multicolumn{1}{c}{\textbf{RNN}} &
  \multicolumn{1}{c}{\textbf{LSTM}} &
  \multicolumn{1}{c}{${\alpha_{t}}$\textbf{-RIM}} & \\
\multicolumn{1}{c}{\textbf{Lag}} &
  \multicolumn{1}{c}{\textbf{MAPE}} &
  \multicolumn{1}{c}{\textbf{MAPE}} &
  \multicolumn{1}{c}{\textbf{MAPE}} & \\ \hline\hline
\textbf{1}        & 2.3279       & 5.2203        & \textbf{1.7081}                 \\
\textbf{2}        & 1.9341       & 5.3022        & 1.7780                  \\
\textbf{3}        & 2.6795       & 5.7144        & 1.9815                 \\
\textbf{4}        & 2.3332       & 6.3815        & 2.1937                 \\
\textbf{5}        & 3.0264       & 5.9082        & 2.3718  \\ 
\bottomrule
\end{tabular}
      \caption{Re-scaled univariate metrics of 10 input lags for the Brown Forman stock.}
    \end{minipage}%
    \hspace{1.6cm}
    \begin{minipage}{.45\linewidth}
      \centering
\begin{tabular}{r r r r r r r} \toprule
\multicolumn{1}{c}{} &
  \multicolumn{1}{c}{\textbf{RNN}} &
  \multicolumn{1}{c}{\textbf{LSTM}} &
  \multicolumn{1}{c}{${\alpha_{t}}$\textbf{-RIM}} & \\
\multicolumn{1}{c}{\textbf{Lag}} &
  \multicolumn{1}{c}{\textbf{MAPE}} &
  \multicolumn{1}{c}{\textbf{MAPE}} &
  \multicolumn{1}{c}{\textbf{MAPE}} & \\ \hline\hline
\textbf{1}        & 4.4243       & 8.7381        & \textbf{1.3157}                 \\
\textbf{2}        & 3.7217       & 9.1994        & 1.6702                 \\
\textbf{3}        & 5.9833       & 9.5561        & 1.9839                 \\
\textbf{4}        & 3.9705       & 9.5775        & 2.2366                 \\
\textbf{5}        & 4.8024       & 9.6431        & 2.4917  \\ 
\bottomrule
\end{tabular}
      \caption{Re-scaled bivariate metrics of 10 input lags for the Brown Forman stock.}
    \end{minipage} 
\end{table}

\vspace{0.3cm}

%############################# 21 input lags Brown Forman #####################################
\textbf{21 Input Lags}\\
\noindent
{\color{lightgray} \rule{\linewidth}{0.2mm} }

\begin{table}[!htb]
\renewcommand*{\arraystretch}{1.2}
    \begin{minipage}{.45\linewidth}
      \centering
        \begin{tabular}{r r r r r r r} \toprule
\multicolumn{1}{c}{} &
  \multicolumn{1}{c}{\textbf{RNN}} &
  \multicolumn{1}{c}{\textbf{LSTM}} &
  \multicolumn{1}{c}{${\alpha_{t}}$\textbf{-RIM}} & \\
\multicolumn{1}{c}{\textbf{Lag}} &
  \multicolumn{1}{c}{\textbf{MAPE}} &
  \multicolumn{1}{c}{\textbf{MAPE}} &
  \multicolumn{1}{c}{\textbf{MAPE}} & \\ \hline\hline
\textbf{1}        & 4.4237       & 7.7809        & \textbf{1.2187}                 \\
\textbf{2}        & 2.9773       & 7.9345        & 1.5905                 \\
\textbf{3}        & 5.4763       & 7.2335        & 1.6570                  \\
\textbf{4}        & 3.1714       & 7.8949        & 1.9313                 \\
\textbf{5}        & 4.1796       & 7.8409        & 2.5372  \\ 
\bottomrule
\end{tabular}
      \caption{Re-scaled univariate metrics of 21 input lags for the Brown Forman stock.}
    \end{minipage}%
    \hspace{1.6cm}
    \begin{minipage}{.45\linewidth}
      \centering
\begin{tabular}{r r r r r r r} \toprule
\multicolumn{1}{c}{} &
  \multicolumn{1}{c}{\textbf{RNN}} &
  \multicolumn{1}{c}{\textbf{LSTM}} &
  \multicolumn{1}{c}{${\alpha_{t}}$\textbf{-RIM}} & \\
\multicolumn{1}{c}{\textbf{Lag}} &
  \multicolumn{1}{c}{\textbf{MAPE}} &
  \multicolumn{1}{c}{\textbf{MAPE}} &
  \multicolumn{1}{c}{\textbf{MAPE}} & \\ \hline\hline
\textbf{1}        & 5.7473       & 11.0067       & \textbf{1.5181}                 \\
\textbf{2}        & 4.2199       & 10.8874       & 2.0879                 \\
\textbf{3}        & 6.0034       & 11.9115       & 2.2657                 \\
\textbf{4}        & 4.8727       & 11.046        & 2.6279                 \\
\textbf{5}        & 4.8366       & 11.172        & 3.0677  \\ 
\bottomrule
\end{tabular}
      \caption{Re-scaled bivariate metrics of 21 input lags for the Brown Forman stock.}
    \end{minipage} 
\end{table}

%=========================== THERMO FISCHER ======================================

\clearpage
\textbf{THERMO FISCHER} \\

%############################# 5 input lags Thermo Fischer #####################################
\textbf{5 Input Lags}\\
\noindent
{\color{lightgray} \rule{\linewidth}{0.2mm} }

\begin{table}[!htb]
\renewcommand*{\arraystretch}{1.2}
    \begin{minipage}{.45\linewidth}
      \centering
        \begin{tabular}{r r r r r r r} \toprule
\multicolumn{1}{c}{} &
  \multicolumn{1}{c}{\textbf{RNN}} &
  \multicolumn{1}{c}{\textbf{LSTM}} &
  \multicolumn{1}{c}{${\alpha_{t}}$\textbf{-RIM}} & \\
\multicolumn{1}{c}{\textbf{Lag}} &
  \multicolumn{1}{c}{\textbf{MAPE}} &
  \multicolumn{1}{c}{\textbf{MAPE}} &
  \multicolumn{1}{c}{\textbf{MAPE}} & \\ \hline\hline
\textbf{1}        & 9.6686       & 11.9074       & \textbf{2.1803}                 \\
\textbf{2}        & 8.3009       & 11.8454       & 2.4475                 \\
\textbf{3}        & 9.9669       & 12.1983       & 2.7177                 \\
\textbf{4}        & 6.0381       & 11.1738       & 2.7458                 \\
\textbf{5}        & 7.5086       & 11.6914       & 3.0844  \\ 
\bottomrule
\end{tabular}
      \caption{Re-scaled univariate metrics of 5 input lags for the Thermo Fischer stock.}
    \end{minipage}%
    \hspace{1.6cm}
    \begin{minipage}{.45\linewidth}
      \centering
\begin{tabular}{r r r r r r r} \toprule
\multicolumn{1}{c}{} &
  \multicolumn{1}{c}{\textbf{RNN}} &
  \multicolumn{1}{c}{\textbf{LSTM}} &
  \multicolumn{1}{c}{${\alpha_{t}}$\textbf{-RIM}} & \\
\multicolumn{1}{c}{\textbf{Lag}} &
  \multicolumn{1}{c}{\textbf{MAPE}} &
  \multicolumn{1}{c}{\textbf{MAPE}} &
  \multicolumn{1}{c}{\textbf{MAPE}} & \\ \hline\hline
\textbf{1}        & 15.4747      & 20.6451       & \textbf{1.8585}                 \\
\textbf{2}        & 11.2671      & 21.233        & 2.2550                  \\
\textbf{3}        & 15.5406      & 21.135        & 2.5692                 \\
\textbf{4}        & 10.8006      & 20.0298       & 2.4522                 \\
\textbf{5}        & 11.6321      & 19.922        & 2.5253  \\ 
\bottomrule
\end{tabular}
      \caption{Re-scaled bivariate metrics of 5 input lags for the Thermo Fischer stock.}
    \end{minipage} 
\end{table}

\vspace{0.3cm}
%############################# 10 input lags Thermo Fischer #####################################
\textbf{10 Input Lags}\\
\noindent
{\color{lightgray} \rule{\linewidth}{0.2mm} }

\begin{table}[!htb]
\renewcommand*{\arraystretch}{1.2}
    \begin{minipage}{.45\linewidth}
      \centering
        \begin{tabular}{r r r r r r r} \toprule
\multicolumn{1}{c}{} &
  \multicolumn{1}{c}{\textbf{RNN}} &
  \multicolumn{1}{c}{\textbf{LSTM}} &
  \multicolumn{1}{c}{${\alpha_{t}}$\textbf{-RIM}} & \\
\multicolumn{1}{c}{\textbf{Lag}} &
  \multicolumn{1}{c}{\textbf{MAPE}} &
  \multicolumn{1}{c}{\textbf{MAPE}} &
  \multicolumn{1}{c}{\textbf{MAPE}} & \\ \hline\hline
\textbf{1}        & 4.9015       & 11.0398       & \textbf{2.7522}                 \\
\textbf{2}        & 3.4432       & 12.4441       & 2.9103                 \\
\textbf{3}        & 5.2922       & 10.1909       & 2.4824                 \\
\textbf{4}        & 3.5335       & 10.8379       & 2.5419                 \\
\textbf{5}        & 4.2818       & 8.9843        & 2.5887  \\ 
\bottomrule
\end{tabular}
      \caption{Re-scaled univariate metrics of 10 input lags for the Thermo Fischer stock.}
    \end{minipage}%
    \hspace{1.6cm}
    \begin{minipage}{.45\linewidth}
      \centering
\begin{tabular}{r r r r r r r} \toprule
\multicolumn{1}{c}{} &
  \multicolumn{1}{c}{\textbf{RNN}} &
  \multicolumn{1}{c}{\textbf{LSTM}} &
  \multicolumn{1}{c}{${\alpha_{t}}$\textbf{-RIM}} & \\
\multicolumn{1}{c}{\textbf{Lag}} &
  \multicolumn{1}{c}{\textbf{MAPE}} &
  \multicolumn{1}{c}{\textbf{MAPE}} &
  \multicolumn{1}{c}{\textbf{MAPE}} & \\ \hline\hline
\textbf{1}        & 17.8271      & 18.6483       & \textbf{1.7189}                 \\
\textbf{2}        & 13.5076      & 19.7892       & 1.9708                 \\
\textbf{3}        & 18.0445      & 20.2516       & 2.2388                 \\
\textbf{4}        & 11.2579      & 18.7996       & 2.4439                 \\
\textbf{5}        & 15.5029      & 17.412        & 2.7314  \\ 
\bottomrule
\end{tabular}
      \caption{Re-scaled bivariate metrics of 10 input lags for the Thermo Fischer stock.}
    \end{minipage} 
\end{table}

\vspace{0.3cm}

%############################# 21 input lags Thermo Fischer #####################################
\textbf{21 Input Lags}\\
\noindent
{\color{lightgray} \rule{\linewidth}{0.2mm} }

\begin{table}[!htb]
\renewcommand*{\arraystretch}{1.2}
    \begin{minipage}{.45\linewidth}
      \centering
        \begin{tabular}{r r r r r r r} \toprule
\multicolumn{1}{c}{} &
  \multicolumn{1}{c}{\textbf{RNN}} &
  \multicolumn{1}{c}{\textbf{LSTM}} &
  \multicolumn{1}{c}{${\alpha_{t}}$\textbf{-RIM}} & \\
\multicolumn{1}{c}{\textbf{Lag}} &
  \multicolumn{1}{c}{\textbf{MAPE}} &
  \multicolumn{1}{c}{\textbf{MAPE}} &
  \multicolumn{1}{c}{\textbf{MAPE}} & \\ \hline\hline
\textbf{1}        & 8.8081       & 9.1374        & 2.8521                 \\
\textbf{2}        & 5.0309       & 9.6582        & 2.7941                 \\
\textbf{3}        & 6.5571       & 9.7462        & \textbf{2.5004}                 \\
\textbf{4}        & 4.1092       & 10.6142       & 2.6365                 \\
\textbf{5}        & 5.0009       & 7.7130         & 2.6647  \\ 
\bottomrule
\end{tabular}
      \caption{Re-scaled univariate metrics of 21 input lags for the Thermo Fischer stock.}
    \end{minipage}%
    \hspace{1.6cm}
    \begin{minipage}{.45\linewidth}
      \centering
\begin{tabular}{r r r r r r r} \toprule
\multicolumn{1}{c}{} &
  \multicolumn{1}{c}{\textbf{RNN}} &
  \multicolumn{1}{c}{\textbf{LSTM}} &
  \multicolumn{1}{c}{${\alpha_{t}}$\textbf{-RIM}} & \\
\multicolumn{1}{c}{\textbf{Lag}} &
  \multicolumn{1}{c}{\textbf{MAPE}} &
  \multicolumn{1}{c}{\textbf{MAPE}} &
  \multicolumn{1}{c}{\textbf{MAPE}} & \\ \hline\hline
\textbf{1}        & 11.537       & 20.6396       & 2.4834                 \\
\textbf{2}        & 8.5560        & 19.5691       & 2.3457                 \\
\textbf{3}        & 14.5108      & 19.4300         & \textbf{2.1511}                 \\
\textbf{4}        & 10.5839      & 19.5961       & 2.4707                 \\
\textbf{5}        & 9.5467       & 19.7083       & 2.8236  \\ 
\bottomrule
\end{tabular}
      \caption{Re-scaled bivariate metrics of 21 input lags for the Thermo Fischer stock.}
    \end{minipage} 
\end{table}
\end{document}